\documentclass[aip,reprint,graphicx]{revtex4-1}
\draft 

\usepackage{multirow}

\usepackage{graphicx}
\usepackage{bm}
\usepackage{amsmath,amssymb}
\usepackage{epstopdf}
\usepackage{mathtools}
\usepackage{epstopdf}
\usepackage{xcolor}
\usepackage{float}

\newcommand{\be}{\begin{equation}}
	\newcommand{\ee}{\end{equation}}

\newcommand{\bea}{\begin{eqnarray}}
	\newcommand{\eea}{\end{eqnarray}}
\newcommand{\beastar}{\begin{eqnarray*}}
	\newcommand{\eeastar}{\end{eqnarray*}}
\newcommand{\nn}{\nonumber\\}

\newcommand{\half}{\frac{1}{2}}

\newcommand{\eq}[1]{~(\ref{#1})}

\newcommand{\order}{{{\mathcal O}}}
\newcommand{\ph}{\check{p}}
\newcommand{\sh}{\check{s}}
\newcommand{\muhs}{\check{\mu}}
\newcommand{\muh}{\bm{\muhs}}
\newcommand{\rh}{\bm{\check{\rho}}}
\newcommand{\rhp}{\rh_+}
\newcommand{\rhm}{\rh_-}
\newcommand{\nh}{\check{n}}
\newcommand{\nhp}{\nh\one}
\newcommand{\nhm}{\nh\two}
\newcommand{\nbar}{\bar{n}}
\newcommand{\f}{\bm{f}}

\newcommand{\ev}{\bm{e}}
\renewcommand{\l}{\bm{l}}
\newcommand{\lt}{\bm{\tilde{l}}}

\renewcommand{\v}{\bm{v}}
\newcommand{\n}{\bm{n}}
\newcommand{\q}{\bm{q}}

\newcommand{\mv}{\bm{m}}
\newcommand{\mvs}{m}
\newcommand{\gv}{\bm{g}}

\renewcommand{\c}{_{\rm c}}
\newcommand{\pt}{\tilde{p}}
\newcommand{\htt}{\tilde{h}}
\renewcommand{\tt}{\tilde{t}}
\newcommand{\rt}{\tilde{\rho}}

\newcommand{\st}{\tilde{s}}

\newcommand{\ltt}{\tilde{l}_2}
\newcommand{\x}{c}

\newcommand{\bt}{\tilde\beta}

\newcommand{\parent}{^{(0)}}
\newcommand{\one}{_+}
\newcommand{\two}{_-}
\newcommand{\onetwo}{_\pm}
\newcommand{\T}{^{\rm T\!}}

\usepackage{array}
\newcolumntype{P}[1]{>{\centering\arraybackslash}p{#1}}
\newcolumntype{M}[1]{>{\centering\arraybackslash}m{#1}}
\newcolumntype{N}{@{}m{0pt}@{}}

\begin{document}


\title{Critical phase behaviour in multi-component fluid mixtures: complete scaling analysis} 



\author{Pablo de Castro}
\email[]{pablo.decastro@kcl.ac.uk}
\affiliation{Disordered Systems Group, Department of Mathematics, King's College London, WC2R 2LS London, United Kingdom}

\author{Peter Sollich}
\email[]{peter.sollich@kcl.ac.uk}
\affiliation{Disordered Systems Group, Department of Mathematics, King's College London, WC2R 2LS London, United Kingdom}
\affiliation{Institut f\"ur Theoretische Physik, Georg-August-Universit\"at, 37077 G\"ottingen, Germany}


\date{\today}

\begin{abstract}
We analyse the critical gas-liquid phase behaviour of arbitrary fluid mixtures in their coexistence region. We focus on the setting relevant for polydisperse colloids, where the overall density and composition of the system are being controlled, in addition to temperature. Our analysis uses the complete scaling formalism and thus includes pressure mixing effects in the mapping from thermodynamic fields to the effective fields of 3D Ising criticality. Because of fractionation, where mixture components are distributed unevenly across coexisting phases, the critical behaviour is remarkably rich. We give scaling laws for a number of important loci in the phase diagram. These include the cloud and shadow curves, which characterise the onset of phase coexistence; a more general set of curves defined by fixing the fractional volumes of the coexisting phases to arbitrary values; and conventional coexistence curves of the densities of coexisting phases for fixed overall density. We identify suitable observables (distinct from the Yang-Yang anomalies discussed in the literature) for detecting pressure mixing effects. Our analytical predictions are checked against numerics using a set of mapping parameters fitted to simulation data for a polydisperse Lennard-Jones fluid, allowing us to highlight crossovers where pressure mixing becomes relevant close to the critical point.
\end{abstract}


\maketitle 


\section{Introduction and Overview}

Soft matter fluids such as colloidal suspensions, polymer solutions, liquid crystals, etc.\ are often composed of non-identical particles, and hence are termed `polydisperse fluids'.\cite{vieville2011polydispersity,auer2001suppression,belli2011polydispersity,sollich2005nematic,ROGOSIC19961337} The polydisperse attribute that distinguishes the different particle species can be particle size, shape, charge, molecular weight, chemical nature, etc.\ \cite{poon2002physics,van2009experimental,stuart1980polydispersity,evans1998universal,PhysRevE.77.011501,PabloPeter1} or a combination of these. (We will mostly use the generic label `size'.)
Polydisperse fluids are very widespread indeed, with examples comprising blood, paint, milk, clay, shampoo, viruses, globular proteins, photonic crystals, pharmaceuticals and even sewage, among several others. \cite{berger1991polydisperse,jose1996maya,anema2008effect,johnson1955particle,lavrinenko2009influence,li2003direct,fraden1995phase,gazzillo2011effects,vera1996colloidal,park2009spectroscopic} The present work investigates how an arbitrary polydisperse fluid in the liquid-gas coexistence region behaves near its liquid-gas critical point (CP). Our development is valid for any number of particle species regardless of the nature of interparticle interactions, and therefore applies to generic multi-component mixtures. We will mostly use the term `polydisperse' for such systems and this should then be read as including both genuinely polydisperse systems such as colloids, where the number of species is effectively infinite, and mixtures with a finite number of components.

Liquid-gas phase-separated polydisperse fluids typically exhibit fractionation, where the overall number of particles from any given species is distributed unevenly across coexisting phases (i.e.\ none of the `daughter' phases has a composition equal to that of the `parent' phase). One therefore expects -- as we will find -- that the critical behaviour of such polydisperse systems is much richer than that of their monodisperse counterparts. For example, consider the case of phase separation starting from a parent phase with a fixed shape of its size distribution. In this scenario a number of different characteristic loci can be defined in the phase diagram. The cloud curve is the one tracing temperature against the overall (parent) density where phase separation first occurs; similarly, the shadow curve records the density of the corresponding incipient phase.\cite{PeterReview} The critical point is located at the intersection of these two curves, rather than at the maximum as in the monodisperse case, where cloud and shadow curves collapse onto a single curve, the standard binodal. If one fixes a parent density one can alternatively study the evolution of the coexisting densities as temperature is lowered, or that of the fractional volumes occupied by the coexisting phases. We will discuss further loci of interest below, in particular ones determined by fixing to arbitrary values the fractions of system volume occupied by the two coexisting phases. This generalises the notion of cloud and shadow curve where these fractions are 1 and 0, respectively.

To motivate the relevance of `complete scaling' (see below) to our analysis, we give an overview of the development of theories for the so-called `diameter' of a fluid. This can be defined, for either monodisperse systems or mixtures, as
\bea
\frac{n_{+}+n_{-}}{2}
\eea
where $n_{\pm}$ are the total number densities of the two coexisting phases obtained on cooling below the critical point. The temperature dependence of this quantity defines a curve in the phase diagram that has historically played an important role in the context of fluid criticality. In the nineteenth century this dependence was described by the `law of rectilinear diameter', i.e.
\bea
\bar{n}\equiv\frac{\nh_{+}+\nh_{-}}{2}\sim |t|
\label{rectilinear}
\eea
This is expressed here in terms of normalised deviations $\nh\equiv (n-n_c)/n_c$ and $t\equiv (T-T_c)/T_c$ from the critical density ($n_c$) and temperature ($T_c$). In the case of monodisperse fluids, such a linear relation between $\bar{n}$ and $t$ can be obtained theoretically via, for example, the van der Waals equation of state. 

However, even for monodisperse fluids, discrepancies between experimental data and the rectilinear law were eventually found in some fluid systems.\cite{PhysRevLett.32.879,shimanskaya1981experimental,PhysRevA.16.2483} A natural hypothesis is that this is due to critical fluctuations, which are treated only approximately in mean-field approaches like van der Waals. To include these, one can construct a scaling formalism that essentially maps three-dimensional Ising critical behaviour onto its fluid counterpart.\cite{PhysRev.87.410,PhysRevLett.13.303,widom1965eqnstate,rehr1973revised,PhysRevE.67.061506} The rationale for such a mapping is the notion of universality generated by the diverging lengthscale of fluctuations at the critical point.
The mapping expresses Ising thermodynamic variables as functions of fluid thermodynamic variables, allowing one to obtain the equation of state of a fluid in the vicinity of a CP. In particular, using the correspondence between the (monodisperse) lattice-gas fluid model and the Ising model, each independent thermodynamic variable in the Ising model is expressed in terms of a single fluid counterpart.\cite{PhysRev.87.410,PhysRevLett.13.303,widom1965eqnstate} This simplest version of a mapping to an Ising scaling theory gives a rectilinear diameter in the sense that a plot of $t$ against $\bar{n}$ is a straight line, though this line is constrained to be vertical ($\bar{n}=0$) because of the particle-vacancy symmetry of the lattice gas.\cite{PhysRevLett.97.025703,PhysRev.87.410}
In order to devise a more flexible mapping framework, Rehr and Mermin introduced `revised' scaling.\cite{rehr1973revised} Here each independent Ising thermodynamic variable (temperature, magnetic field) is a function of \textit{all independent} fluid thermodynamic variables (temperature, chemical potential): the thermodynamic fields have been `mixed', producing `asymmetric' fluid criticality. The leading-order behaviour for the diameter in this case comes out as $\bar{n}\sim |t|^{1-\alpha}$, where $\alpha$ is the universal critical exponent of the specific heat. This prediction agrees with experimental data for some fluid systems but still leaves out a number of other situations.\cite{PhysRevB.36.599} It was only after the beginning of the current century that the formalism known now as `complete' scaling was introduced,\cite{PhysRevE.67.061506} whereby in the critical region each Ising scaling variable -- temperature, magnetic field and (the singular part of the) thermodynamic potential -- is expressed as a combination of \textit{all} relevant thermodynamic variables of the fluid (temperature, chemical potential and pressure). This introduces `pressure mixing', in the sense that the fluid pressure now appears in the expressions for all Ising thermodynamic variables.  For the diameter, the complete scaling approach predicts, in addition to the $1-\alpha$ term, a new, more singular contribution, namely $\bar{n}\sim |t|^{2\beta}$ where $\beta$ is the critical exponent for the spontaneous magnetisation. There are experimental data for monodisperse fluid systems which support this prediction.\cite{PhysRevLett.97.025703}

Since then a number of complete scaling studies have been produced, but only a few of them have looked at the behaviour of fluid \textit{mixtures}. This is an important gap for soft matter, where multi-component systems are much more common than one-component ones.\cite{sollich2011polydispersity,sluckin1989polydispersity,liddle2011polydispersity} In Refs.\ \citenum{ISI:000254539700040} and \citenum{ISI:000276971500024} the case of a binary mixture is considered, but there the fluid pressure is controlled; in Ref.\ \citenum{ISI:000254539700040} this is done by considering an incompressible fluid mixture setup, whereas in Ref.\ \citenum{ISI:000276971500024} pressure effects are included only indirectly via their effect on mapping coefficients, in such a way that the pressure is effectively fixed. The scaling behaviour one then predicts is essentially that of a monodisperse system, though with quantitative changes in coefficients that effect e.g.\ in which direction the diameter curves.\cite{ISI:000276971500024}

Here we develop a complete scaling framework for generic multi-component fluids (not only binary mixtures) where the overall number density of particles is fixed rather than allowed to fluctuate at fixed pressure; the overall composition (fraction of particles belonging to each species) is also fixed. This is the natural setting for colloids and other soft matter fluids, where density is easily fixed by the amount of dilution using a solvent. Controlled pressure, which is common for atomic and molecular fluids, would correspond to the more unusual situation of fixed {\em osmotic} pressure for colloids. In addition to obtaining new results for the `diameter', we look at a comprehensive set of other phase-diagram curves. In Refs.\ \citenum{Belyakov2013} and \citenum{ISI:000345509900008}, Belyakov \textit{et al.}\ developed a similar framework for multi-component systems with fixed composition, but they employed their results to create a fitting technique for experimental data near the critical point rather than investigating as we do the scaling exponents of the various characteristic curves in the phase diagram of a mixture. (Note also that in Ref.\ \citenum{Belyakov2013} the effects of complete scaling were not included explicitly, although as we show below this does not change the qualitative scaling form of the cloud curve.)

We will consider in this paper a generalisation of cloud and shadow curves that reveals interesting structures inside the coexistence region. To motivate this, note first that in the monodisperse case one can define the diameter in at least two distinct, \textit{equivalent} ways. The usual one is to define the diameter as the temperature-dependent midpoint of the phase coexistence region, i.e.\ $(\nh_{+}+\nh_{-})/2$. The coexisting densities can be generated by cooling a system with the critical parent density $n_c$.
Alternatively, one can define the diameter as the parent density that produces, at each temperature, two phases each occupying half the system volume: due to the lever rule, which expresses particle conservation, this parent density must then be the average of the two coexisting densities.

In the polydisperse case, we will see below that these two definitions are \textit{not} equivalent, due to fractionation effects. We will refer to the first construction as `midpoint diameter', defined as the average density of coexisting phases obtained by cooling a critical parent system. For the second definition we will use the term `equal volume diameter'. Experimentally this curve could be obtained by fixing density and decreasing temperature from outside the coexistence region: in this process one crosses the cloud curve, where the split of fractional volumes between the coexisting phases is $100$--$0\%$, and then has to cool further until the split becomes $50$--$50\%$. Generalising this construction, we will consider below the `fixed fractional volume' loci in the phase diagram where the parent density is chosen at each temperature such as to produce a {\em fixed} split of fractional volumes, say $80$--$20\%$, between the coexisting phases. We will see that the two different definitions of the diameter and the fixed fractional volume lines, which to our knowledge have not been analysed before, provide useful probes of fluid mixture critical behaviour.

In summary, in this work we use a complete scaling theory to relate polydisperse criticality to standard 3D Ising criticality, and thus to predict the scaling of the various lines in the phase diagram. We will investigate which nonlinear field mixing terms need to be retained to account properly for fractionation effects, and will point out observables, distinct from the Yang-Yang anomalies used previously,\cite{PhysRevE.67.061506} that could be useful to detect potential pressure mixing effects.

The remainder of this article is structured as follows. In section \ref{model} we set up the complete scaling theory and derive the key relations between the thermodynamic variables. In section \ref{coexistencecond} we work out the general conditions from which the properties of coexisting phases can be determined. Section \ref{fractional} contains our discussion of constant fractional volume lines. These include the cloud curve and, via the density of the coexisting phases, also the shadow curve. In section \ref{mono} we pause briefly to discuss in which limit our results reduce to the monodisperse case and how this leads to qualitatively different scaling behaviour. Resuming the discussion of characteristic loci in the phase diagram, the scaling behaviour of the (conventional, fixed parent density) coexistence curves is discussed in section \ref{coexistence}. Then in Section \ref{numerics} we verify our analytical derivations by numerically solving the complete scaling equations, for a set of mapping coefficients that reproduces cloud and shadow curve data obtained in Monte Carlo simulations of a polydisperse Lennard-Jones fluid.\cite{LJNigelPeter} A summary, conclusions, and discussion of our results can be found in section \ref{conc}. The appendices contain technical details of the calculations required to extract the various scaling laws and to establish the correspondence with the monodisperse limit, as well as information on how we fitted the complete scaling model to simulation data.

\section{Complete scaling setup}
\label{model}

In the critical region the thermodynamic behaviour of Ising-like systems is governed by two independent Ising-like scaling fields: a temperature variable (or thermal field) denoted by $\tt$ and a field variable (or ordering field) denoted by $\htt$. These two variables determine the singular part of an appropriate pressure-like variable (i.e.\ the Ising thermodynamic potential) $\pt$. We assume these three variables are defined such that they are zero at criticality. Asymptotically close to the critical point, $\pt$ becomes a generalised homogeneous function of $\tt$ and $\htt$ of the form
\be
\pt = Q|\tt|^{2-\alpha}f^{\pm}\left(\frac{\htt}{|\tt|^{2-\alpha-\beta}}\right)
\label{pt_scaling_form}
\ee
where $Q$ is a positive amplitude, $f^{\pm}$ are two universal scaling functions that encode the properties of the three-dimensional Ising universality class, with the superscripts $\pm$ indicating $\tt \gtrless 0$. 

 In Ref.\ \citenum{PhysRevE.67.061506}, Kim, Fisher, and Orkoulas (hereafter KFO) introduced the complete scaling approach, in the context of one-component fluids. In this formalism each of the Ising-model variables ($\pt$, $\tt$, and $\htt$) is expressed as a function of \textit{all} thermodynamic fluid variables (pressure, temperature, and chemical potential), and this allows one to use Ising relations to work out the scaling behaviour of the fluid. (Formally this approach stems from a principle of isomorphism.\cite{ISI:000254539700040}) We now extend the approach to the case of a polydisperse fluid, similarly to Ref.\ \citenum{ISI:000345509900008}. To do so we proceed by writing second-order expansions around the critical point for the Ising scaling variables in terms of the fluid variables (now pressure, temperature, and \textit{species chemical potentials}) as
\bea
\pt = {}
&\ph - k_0 t - \l_0\T\muh - r_0 t^2 - \muh\T\q_0\muh \nn
&- t \v_0\T\muh - m_0\ph^2 - n_0 \ph t - \ph \n_3\T\muh
\label{pt_mix}\\
\tt ={} 
& t - \l_1\T\muh -j_1 \ph - r_1 t^2 - \muh\T\q_1\muh \nn
&- t \v_1\T\muh - m_1\ph^2 - n_1 \ph t - \ph \n_4\T\muh
\label{tt_mix}\\
\htt ={} 
&\l_2\T\muh - k_2 t -j_2 \ph - r_2 t^2 - \muh\T\q_2\muh \nn
&- t \v_2\T\muh - m_2\ph^2 - n_2 \ph t - \ph \n_5\T\muh
\label{ht_mix}
\eea
where the entire set of coefficients $\l_i,k_i,j_i,r_i,\q_i,\v_i,m_i$, etc.\ appearing in Eqs.\ (\ref{pt_mix})--(\ref{ht_mix}) are dubbed `mixing coefficients' and
\be
t \equiv \frac{T-T\c}{T\c}, \qquad
\ph \equiv \frac{p-p\c}{n\c k_B T\c}
\ee
Here $n\c$ and $T\c$, the critical density and temperature (for a fixed parent composition or
`dilution line'\cite{PeterReview}) are used to make all quantities dimensionless; the critical pressure is denoted by $p\c$ and $k_B$ is Boltzmann's constant. The vector $\muh$ has as many components as there are species in the fluid, each of them being denoted by
\be
\muhs(\sigma) \equiv \frac{\mu(\sigma)-\mu\c(\sigma)}{k_B T\c}
\ee
where $\mu(\sigma)$ is the chemical potential of a species labelled by an arbitrary polydisperse attribute $\sigma$ and $\mu\c(\sigma)$ is its critical value. Because $\muh$ is a vector, the quadratic expansions in Eqs.\ (\ref{pt_mix})--(\ref{ht_mix}) require appropriate `mixing coefficients' vectors and matrices when $\muh$ appears (instead of the scalar constants in the other cases); all vectors are taken as column vectors and $\ldots\T$ denotes the transpose of a vector. The vector $\l_2$ replaces a unit constant in KFO; leaving this
unconstrained means no extra scaling of the argument of $f^\pm$
is needed. We use the notation $k_2$ instead of the $k_1$ in KFO to ensure that the subscript of each mixing coefficient in Eqs.\ (\ref{pt_mix})--(\ref{ht_mix}) specifies uniquely to which of the Ising scaling variables $\pt$, $\tt$, and $\htt$ it belongs.

Denoting the arbitrary number of species in the fluid by $M$, we can check that the above construction gives the right number of equations for a proper equation of state: we have $M+2$ fluid thermodynamic variables ($\ph$, $t$, and $\muh$), $3$ Ising scaling variables ($\pt$, $\tt$, and $\htt$), and $3+1$ equations [Eqs.\ (\ref{pt_scaling_form})--(\ref{ht_mix})]. Specifying $M+1$ thermodynamic variables ($t$ and $\muh$) then determines all other variables and so in particular the pressure $\ph$.

At phase coexistence, $t$, $\muh$ and $\ph$ are the same in both
phases. Hence, from the above expansions [Eqs.\ (\ref{pt_mix})--(\ref{ht_mix})], so are $\tt$, $\htt$ and $\pt$.
Thus the relation between these Ising scaling variables along the phase boundary can be worked out from the universal scaling function. Conventionally, one would parametrise this
dependence by $\tt$ ($<0$); then $\htt$, $\pt$, $\rt$, $\st$ are, at
least in principle, known functions of $\tt$, where the generalised number density, $\rt$, and entropy density, $\st$, are the Ising scaling densities, defined by the relation $d\pt =\rt\,d\htt+\st\,d\tt$. The results worked out by KFO (omitting prefactors and using the subscripts $\pm$ to label now the two phases) can be written as follows:
\bea
\pt&\sim&|\tt|^{2-\alpha}+\ldots\\
\htt&\sim& |\tt|^{2-\alpha-\beta+\theta^\prime}+\ldots
\label{ht_scaling}
\\
\rt\onetwo&\sim& \pm\left(|\tt|^\beta + |\tt|^{\beta+\theta}\pm
|\tt|^{\beta+\theta^\prime}\right)
\label{rt_scaling}
\\
\st\onetwo&\sim& |\tt|^{1-\alpha} + |\tt|^{1-\alpha+\theta} \pm
|\tt|^{1-\alpha+\theta^\prime}.
\label{st_scaling}
\eea
For the relevant Ising 3D case, KFO quote $\beta\simeq0.326$,
$\alpha\simeq0.109$, $\theta\equiv \theta_4\simeq0.52$, $\theta^\prime\equiv\theta_5\simeq1.32$; the latter two are exponents for the leading (even/odd) corrections to scaling. (KFO point out that in contrast to the symmetric case, $\tilde{h}$ does not vanish identically along the phase boundary.)

We will not keep track of terms of order higher than $|\tt|^1$. Therefore, we will neglect $\pt$ and $\htt$. In the scaling (\ref{rt_scaling}) for $\rt$, one in principle needs the
leading correction to scaling, but one can avoid having to take this
into account explicitly by parametrising everything in terms of
$\rt\one\equiv\rt$, which we assume to be positive.
In particular, one has
\be
\tt =  -a\rt^{1/\beta}+\ldots\\
\label{tt_vs_rt}
\ee
with an appropriate constant $a$, and for the entropy density, by
eliminating $\tt$ from the scalings\eq{rt_scaling} and\eq{st_scaling},
\be
\st \sim  \rt^{(1-\alpha)/\beta} + \rt^{(1-\alpha+\theta)/\beta}
\label{st_vs_rt}
\ee
The first exponent is $1/\bt \simeq 2.73$, the inverse of the
`Fisher-renormalised' order parameter exponent $\bt=\beta/(1-\alpha)$,
while the second one, from the corrections to scaling, is
greater than $1/\beta\simeq3.07$. Since we are only keeping terms to
$\rt^{1/\beta}$ [see Eq.\ (\ref{tt_vs_rt})], we need only to retain the first term in scaling (\ref{st_vs_rt}). Thus one can write
\be
\st = -b\rt^{1/\bt} + \ldots
\label{expansionentropy}
\ee
where $b$ is some constant.

As a technical aside, since we will frequently need to invert several series expansions
with non-integer exponents, it is useful to recall that the inverse
series of
\be
y=\sum_{i=0}^\infty a_i x^{n_i}
\label{GeneralInversion}
\ee
(with increasing exponents $0<n_0<n_1<\ldots$) has the form
\be
x = \left(\frac{y}{a_0}\right)^{1/n_0}\left(1+\sum_{i\geq 1}b_i y^{n'_i}
+\sum_{i,j\geq 1}b_{ij} y^{n'_i+n'_j} + \ldots\right)
\ee
with $n_i'\equiv(n_i-n_0)/n_0$ and this is defined only when $y$ and $a_0$ have the same sign, a restriction that we will mostly omit in similar results below. Applied to the scaling\eq{rt_scaling} this gives
\be
\tt \sim \rt^{1/\beta}\left(1+\rt^{\theta/\beta}+\rt^{\theta^\prime/\beta}+\ldots\right)
\ee
and hence the expansion\eq{st_vs_rt}. Note that in Eq.\ (\ref{expansionentropy}) the leading-order term $-b\rt^{1/\bt}$ that we are keeping is identical for the two
coexisting phases, and hence we have dropped the $\pm$ subscript from $\st$. Likewise,
$\rt\two = -\rt + \order(\rt^{1+\theta^\prime/\beta})\approx -\rt$, to the
order of our expansion. Overall, we see that deviations from the Ising
symmetry would only make themselves felt at higher orders.

We will need to know, for a given phase of the system, the density distribution vector $\rh$. Its components are the normalised species densities $\check{\rho}(\sigma)\equiv\rho(\sigma)/n\c$.
This vector of densities can be found by taking the
$\muh$-derivative of the pressure $\ph$; the analogous derivative
with respect to temperature $t$ is the entropy density $\sh=s/n\c$.  The Ising scaling analogues of these quantities are $\st=\partial\pt/\partial\tt$ and $\rt=\partial\pt/\partial\htt$, respectively. Since  $d\pt = \st\,d\tt + \rt\,d\htt$, one can write, by analogy to the derivation in KFO,
\bea
d\pt &=& \st\left(
\frac{\partial\tt}{\partial\ph}d\ph + 
\frac{\partial\tt}{\partial  t}d  t +
\frac{\partial\tt}{\partial\muh}d\muh\right)
\nn
& &{} +\rt \left(
\frac{\partial\htt}{\partial\ph}d\ph + 
\frac{\partial\htt}{\partial  t}d  t +
\frac{\partial\htt}{\partial\muh}d\muh\right)
\\
&=& \frac{\partial\pt}{\partial\ph}d\ph + 
\frac{\partial\pt}{\partial  t}d  t +
\frac{\partial\pt}{\partial\muh}d\muh.
\eea
Inserting the Gibbs\textendash Duhem equation $d\ph=\sh dt + \rh \T d\muh$ (and imposing equality of the
coefficients of $d\muh$) gives
\be
\rh = \frac{
	-   \frac{\partial\pt}{\partial\muh} + 
	\st \frac{\partial\tt}{\partial\muh} +
	\rt \frac{\partial\htt}{\partial\muh}}
{   \frac{\partial\pt}{\partial\ph} -
	\st\frac{\partial\tt}{\partial\ph} -
	\rt\frac{\partial\htt}{\partial\ph}}.
\label{rh_gen}
\ee
This is the mixture analogue of Eq.\ (5) in Ref.~\citenum{ISI:000345509900008}. Inserting the above expansions [Eqs.\ (\ref{pt_mix})--(\ref{ht_mix})] into Eq.\ (\ref{rh_gen}) and re-expanding (in terms of $\ph$, $t$, $\muh$, $\rt$, and $\st$) leads to
\bea
\rh\onetwo &=& \l_0 + (2m_0\l_0+\n_3)\ph + (n_0\l_0+\v_0)t
\nn
& &{}+ (\l_0\n_3\T+2\q_0)\muh+ \order_2+\order_3\pm\rt(\lt_2+\order_1+\order_2) 
\nn
& &{} +\rt^2(- j_2\lt_2+\order_1)\pm\rt^3 j_2^2\lt_2+\st(-j_1\l_0-\l_1)+ \ldots
\nn
\label{rh_mix}
\eea
where $\lt_2\equiv\l_2-j_2\l_0$ and we have specialised to the two coexisting
phases. In writing down Eq.\ (\ref{rh_mix}) we have anticipated that $\ph$, $t$
and $\muh$ will be no larger than $\sim \rt$ and have thrown away
contributions which as a result are smaller than $\rt^{1/\beta}$, e.g.\ terms
$\sim \rt^4$, $\rt\st$, $\st^2$, and so on. The $\order_{1,2,3}$
symbols represent terms in $(\ph,t,\muh)$ of the order indicated and
will either not be crucial below or cannot be written down explicitly
without including third or higher order terms in the mapping expansions 
Eqs.\ (\ref{pt_mix})--(\ref{ht_mix}).

We notice by inspecting Eq.\ (\ref{rh_mix}) that, if only the linear coefficients ($j_i$, $k_i$, and $\l_i$) are included, then $\rh\onetwo$ is a vectorial combination of $\l_0$, $\l_1$, and $\l_2$. Therefore for $M>3$ it becomes impossible to realize a generic size distribution in the coexisting phases. Thus we conclude that in order to have fractionation properly accounted for one needs to include nonlinear mixing coefficients. 

The critical scaling between the Ising variables has now been written down and we have expressed the density distribution vectors of the polydisperse fluid in terms of its thermodynamic variables and of the Ising scaling densities. It remains to add the particle conservation condition for each species, which we do in the next section. Putting everything together one has a system of equations that can be solved either numerically or analytically by expansion, allowing one to obtain the critical phase behaviour in terms of the physical fluid variables only.

\section{Coexistence conditions}
\label{coexistencecond}

One of our goals is to obtain the critical scaling versions of cloud and shadow curves and, more generally, information on coexisting phases. The conditions of equal $\ph$, $t$ and $\muh$ will be satisfied if we are somewhere on the scaling phase boundary, as
parametrised by $\rt$. In addition, we need to satisfy particle
conservation, i.e.\ the `dilution line' constraint. Let us write the parent
density distribution as $\rh\parent=(1+\nh)\f$, where the `$\left(0\right)$' superscript indicates the parent phase, $\f$ is the
normalised parent density distribution vector and $\nh \equiv (n\parent-n\c)/n\c$ is
the (normalised) deviation of the parent density from its critical
value. We write the fractional phase volumes as $\half(1\pm\Delta)$ so that $\Delta=0$ represents the situation where both phases occupy equal volumes.
Then we need to satisfy 
\bea
(1+\nh)\f = \half(1+\Delta)\rhp +
\half(1-\Delta)\rhm
\label{dilution}
\eea
where $\rhp$ and $\rhm$ are given by Eq.~(\ref{rh_gen}) with $\tilde\rho_\pm=\pm\tilde \rho$ inserted. For our scaling expansions we use the expanded form\ (\ref{rh_mix}) instead of Eq.~(\ref{rh_gen}), leading to
\bea
(1+\nh)\f&=&\half(1+\Delta)[
\l_0 + (2m_0\l_0+\n_3)\ph + (n_0\l_0+\v_0)t 
\nn
& &{} + (\l_0\n_3\T+2\q_0)\muh
+\order_2+\order_3 + \rt(\lt_2+\order_1
\nn
& &{}+\order_2)+\rt^2(- j_2\lt_2+\order_1)+\rt^3j_2^2\lt_2 +\st(-j_1\l_0
\nn
& &{} -\l_1)]+\half(1-\Delta)[
\l_0 + (2m_0\l_0+\n_3)\ph 
\nn
&&{}+ (n_0\l_0+\v_0)t + (\l_0\n_3\T+2\q_0)\muh+\order_2+\order_3 
\nn
&&{}- \rt(\lt_2+\order_1+\order_2)+\rt^2(- j_2\lt_2+\order_1)
\nn
&&{}-\rt^3j_2^2\lt_2 +\st(-j_1\l_0-\l_1)].
\label{1plusnh}
\eea
At the critical point, both density distributions must be equal to the
parent one. By setting all thermodynamic variables to zero in Eq.\ (\ref{1plusnh}), one can see that $\l_0=\f$; Eq.\ (\ref{1plusnh}) can in turn be simplified to
\bea
\nh\f &=& (2m_0\f+\n_3)\ph + (n_0\f+\v_0)t + (\f\n_3\T+2\q_0)\muh
\nn
& &{}+\rt^2(- j_2\lt_2+\order_1) + \order_2+\order_3+\st(-j_1\f-\l_1)
\nn
&&{}+\Delta\rt(\lt_2+\order_1+\order_2)
+\Delta\rt^3\,j_2^2\lt_2.
\label{coex_cond}
\eea

We can now write down the set of equations that we need to solve, bearing
in mind that $\pt$ and $\htt$ have been neglected to our order of
expansion and noting explicitly the omitted third-order terms in the
mapping expansions [Eqs.\ (\ref{pt_mix})--(\ref{ht_mix})]:
\bea
0   &=& \ph - k_0 t - \f\T\muh + \order_2 + \order_3
\label{cond0_mix}
\\
\tt   &=& t - \l_1\T\muh - j_1 \ph + \order_2 + \order_3
\label{cond1_mix}
\\
0   &=& \l_2\T\muh - k_2 t - j_2 \ph + \order_2 + \order_3
\label{cond2_mix}
\eea
and
\bea
&-&\Delta\rt\lt_2
+ j_2\lt_2\rt^2-\Delta\rt^3\,j_2^2\lt_2
+ \st(j_1\f+\l_1) 
= -\nh\f 
\nn
&+& (2m_0\f+\n_3)\ph + (n_0\f+\v_0)t + (\f\n_3\T+2\q_0)\muh 
\nn
&+& \order_2 + \order_3 + \Delta \rt(\order_1+\order_2) + \rt^2\,\order_1.
\label{cond3_mix}
\eea
One sees that in this approach, fixed fractional volume lines appear naturally as they correspond to fixed $\Delta$. These lines can then be traced out by considering a series of increasing $\rt$ (or corresponding $\tt$) and for each $\rt$ solving the above $M+3$ equations for the $M+3$ unknowns ($\nh$, $\ph$, $t$, $\muh$). For the cloud curve one would fix $\Delta=1$ for the high density
branch and $\Delta=-1$ for the low-density branch (where the high density phase is the shadow phase and occupies a vanishing fraction of the system volume).
These cases with constant $\Delta$ are considered in Section \ref{fractional}. For actual coexistence curves (see Section \ref{coexistence}) one wants to fix the parent density $\nh$ instead and
infer $\Delta$. This can be done by treating $\Delta\rt$ as a small quantity
to expand in, in addition to $\rt$; note that $\Delta\rt$ can be
much smaller than $\rt$ but no larger since $|\Delta|\leq 1$. We then
have to eliminate $\Delta\rt$ in the end by using the constraint of
fixed $\nh$.

From the structure of the conditions above one sees that $\nh$, $\ph$, $t$
and $\muh$ will be smooth functions of the `inputs' on the left-hand sides of Eqs.\ (\ref{cond0_mix})--(\ref{cond3_mix}), i.e.\ $\Delta\rt$, $\rt^2$, $\st=-b\rt^{1/\bt}$,
and $\tt=-a\rt^{1/\beta}$; the input term $\Delta\rt^3$ is already covered here as the product of $\Delta\rt$ and $\rt^2$. The input variables appearing on the right-hand side
in Eq.\ (\ref{cond3_mix}) have quantitative effects, but do not produce new
terms in the expansion. Thus we can write
\bea
\nh &=& \nu_1\Delta\rt + (\nu_2+\nu_2'\Delta^2)\rt^2 + \nu_3 \rt^{1/\bt}
\nn
&&{}
+ (\nu_4\Delta+\nu_4'\Delta^3)\rt^3+ \nu_5 \rt^{1/\beta}
\label{nh_exp}
\\
\ph &=& \pi_1\Delta\rt + (\pi_2+\pi_2'\Delta^2)\rt^2 + \pi_3 \rt^{1/\bt}
\nn
&&{}+ (\pi_4\Delta+\pi_4'\Delta^3)\rt^3 + \pi_5 \rt^{1/\beta}
\label{ph_exp}
\\
t   &=& \tau_1\Delta\rt + (\tau_2+\tau_2'\Delta^2)\rt^2 + \tau_3 \rt^{1/\bt} 
\nn
&&{}+ (\tau_4\Delta+\tau_4'\Delta^3)\rt^3+ \tau_5 \rt^{1/\beta}
\label{t_exp}
\\
\muh&=& \mv_1\Delta\rt + (\mv_2+\mv_2'\Delta^2)\rt^2 + \mv_3 \rt^{1/\bt}
\nn
&&{}+ (\mv_4\Delta+\mv_4'\Delta^3)\rt^3 + \mv_5 \rt^{1/\beta}
\label{muh_exp}
\eea
where we have introduced appropriate coefficients $\nu_i, \nu^\prime_i, \pi_i, \pi^\prime_i, \tau_i, \tau^\prime_i, \mv_i$, and $\mv^\prime_i$, with the latter two types representing vectors of coefficients with $M$ components each. Notice that as throughout, we do not keep track of terms of order higher than $\rt^{1/\beta}$ here.

Pressure mixing coefficients are defined as the coefficients of the terms where the fluid pressure variable $\ph$ appears in the expansions for $\tt$ and $\htt$, i.e.\ Eqs.\ (\ref{tt_mix}) and (\ref{ht_mix}). (We exclude from this definition the coefficients in the expansion (\ref{pt_mix}) for the pressure-like Ising variable $\pt$.)
Without pressure mixing, where the $\rt^2$ and $\Delta \rt^3$ terms
on the left-hand side of Eq.\ (\ref{cond3_mix}) are absent ($j_2=0$ in this case), one has an expansion in $\Delta\rt$,
$\st$ and $\tt$ only, so that $\nu_2, \pi_2, \tau_2, \mv_2$ and $\nu_4,
\pi_4, \tau_4, \mv_4$ all vanish. One can check that the $\rt^2\,\order_1$ term on the right-hand side of Eq.\ (\ref{cond3_mix}) does not affect this conclusion.

By inserting Eqs.\ (\ref{nh_exp})--(\ref{muh_exp}) into Eqs.\ (\ref{cond0_mix})--(\ref{cond3_mix}), and comparing terms order by order,
we obtain sets of equations that involve no thermodynamic variables, i.e.\ they contain only coefficients. These can be solved for the coefficients $\nu_i, \nu^\prime_i, \pi_i, \pi^\prime_i, \tau_i, \tau^\prime_i, \mv_i$, and $\mv^\prime_i$, in terms of the `mixing coefficients'. (See Appendix \ref{CoeffsEqns}.)

\subsection{Coexisting density distributions}

As part of the output of the calculation one wants to look at the
coexisting density distributions.
Comparing Eq.\ (\ref{rh_mix}) with Eq.\ (\ref{coex_cond}) shows that
\be
\rh\onetwo - \f = \nh\f +(\pm
1-\Delta)[\rt(\lt_2+\order_1+\order_2)+\rt^3 j_2^2\lt_2].
\ee
Once we insert the expansions of $\ph$, $t$, $\muh$ [Eqs.\ (\ref{ph_exp})--(\ref{muh_exp})] into the
$\order_1$ and $\order_2$ terms we see that they contribute with terms
scaling as $\Delta\rt$, $\rt^2$ (except if there is no pressure
mixing) and $\Delta^2\rt^2$, so that
\be
\rh\onetwo - \f = \nh\f +(\pm
1-\Delta)[\rt\lt_2+\gv_2\Delta\rt^2+\gv_4\rt^3+\gv_4'\Delta^2\rt^3]
\ee
with some vectors $\gv_2$, $\gv_4$ and $\gv_4'$; $\gv_4$ vanishes
without pressure mixing. Along with $\lt_2$,
these determine the directions in density distribution space along which
fractionation takes place for the given parent composition $\f$; to linear
and quadratic orders in $\rt$, there is one such direction each, and two to third order.

For the overall coexisting densities themselves one has, by taking the
product with $\ev\T$ where $\ev$ is a vector with all components equal to $1$, the following expression:
\bea
\nh_\pm &=& 
\nh + (\pm1-\Delta)\Bigl[\ltt\rt + g_2\Delta\rt^2+g_4\rt^3
\nn
&&{}+g_4'\Delta^2\rt^3\Bigr]
\label{nh_pm2}
\\
&=& 
\rt\left[\pm \ltt +
\Delta(\nu_1-\ltt)\right]
+\rt^2\Bigl[\nu_2\pm g_2\Delta
\nn
& &{}+(\nu_2'-g_2)\Delta^2\Bigr]+\nu_3\rt^{1/\bt}+\rt^3\Bigl[\pm g_4
\nn
& &{}
+(\nu_4-g_4)\Delta \pm g_4'\Delta^2+(\nu_4'-g_4')\Delta^3\Bigr]
\nn
& &{}
+\nu_5\rt^{1/\beta}.
\label{nh_pm}
\eea
Here we have abbreviated the element sums (not norms!) of the various vectors as $\ltt \equiv \ev\T\lt_2$, $g_2\equiv\ev\T\gv_2$ etc. In the second step we have inserted Eq.\ (\ref{nh_exp}). Note 
that one expects $\ltt>0$ in order to ensure $\nhp>\nhm$; also,
for $\Delta=1$ the (liquid cloud) parent density should increase with
$\rt$, so that $\nu_1$ should likewise come out positive.

\section{Constant fractional volume lines}
\label{fractional}

We can now look at the various curves in the phase diagram that are
obtained for fixed $\Delta$. With $\Delta$ fixed, there is only $\rt$
to eliminate as the curve parameter since $\Delta \rt$ is no longer treated as an additional small quantity to expand in. The elimination process involves
inverting expansions like Eq.\ (\ref{nh_exp}), which leads to
\bea
\rt &=& \frac{\nh}{\nu_1\Delta}
- \frac{\nu_2+\nu_2'\Delta^2}{\nu_1\Delta}\left(\frac{\nh}{\nu_1\Delta}\right)^2
- \frac{\nu_3}{\nu_1\Delta}
\left(\frac{\nh}{\nu_1\Delta}\right)^{1/\bt}
\nn
& &{}+
\left[2\left(\frac{\nu_2+\nu_2'\Delta^2}{\nu_1\Delta}\right)^2
-\frac{\nu_4+\nu_4'\Delta^2}{\nu_1\Delta}\right]
\left(\frac{\nh}{\nu_1\Delta}\right)^3
\nn
&&{}-\frac{\nu_5}{\nu_1\Delta}
\left(\frac{\nh}{\nu_1\Delta}\right)^{1/\beta}.
\label{rt_inversion}
\eea
As pointed out after Eq.\ \eqref{GeneralInversion}, our convention here and below is that expansions involving non-integer powers without modulus signs should be interpreted as applicable only when the quantity being raised is positive. Eq.\ (\ref{rt_inversion}) holds for $\Delta\neq 0$, while for $\Delta=0$ one has similarly
\bea
\rt &=& \left(\frac{\nh}{\nu_2}\right)^{1/2} -
\frac{\nu_3}{2\nu_2}\left(\frac{\nh}{\nu_2}\right)^{(1/\bt-1)/2}
\nn
& &{}
-\frac{\nu_5}{2\nu_2}
\left(\frac{\nh}{\nu_2}\right)^{(1/\beta-1)/2}.
\label{rt_inversion0}
\eea
In both cases the expansions are given to the order that can be
determined reliably from the original expansions up to $\rt^{1/\beta}$.

Inserting Eq.\ (\ref{rt_inversion}) into Eq.\ (\ref{t_exp}) yields for the
temperature as a function of the parent density (when $\Delta\neq 0$):
\bea
t &=& \frac{\tau_1}{\nu_1} \nh
- \tilde\tau_2'
\left(\frac{\nh}{\nu_1}\right)^2
- \tilde\tau_3\left(\frac{\nh}{\nu_1\Delta}\right)^{1/\bt}
\nn
& &{}-
\biggl[\tilde\tau_4\Delta+\tilde\tau_4'\Delta^3\biggr.
\biggl.-2\left(\frac{\nu_2+\nu_2'\Delta^2}{\nu_1\Delta}\right)
\tilde\tau_2'\Delta^2\biggr]
\left(\frac{\nh}{\nu_1\Delta}\right)^3
\nn
&&{}
- \tilde\tau_5
\left(\frac{\nh}{\nu_1\Delta}\right)^{1/\beta}
\label{const_delta}
\eea
where
\be
\tilde\tau_i\equiv \nu_i\frac{\tau_1}{\nu_1} -\tau_i, \qquad
\tilde\tau_i'\equiv \nu_i'\frac{\tau_1}{\nu_1} -\tau_i'
\ee
Note that with this definition one has $\tilde\tau_2=0$, a fact we have already used above. (This result comes from a general proportionality between first and second order expansion coefficients, $\nu_2=-j_2\nu_1$, $\tau_2=-j_2\tau_1$
etc, which we derive in Appendix \ref{CoeffsEqns}.)
The coefficient structure makes
sense: in the hypothetical degenerate case where the $\rt$-expansion
coefficients for $t$ [Eq.~\eqref{t_exp}] and $\nh$ [Eq.~\eqref{nh_exp}] were all proportional to each other,
then $t$ and $\nh$ themselves would be proportional and therefore all terms must vanish (as is ensured by the definition of the $\tilde\tau_i$) except for $t=(\tau_1/\nu_1)\nh$.

As indicated, the linear term in Eq.~\eqref{const_delta} is {\em independent} of $\Delta$; the only $\Delta$-dependence arises via the higher order
terms, and for the singular contributions it is a simple scaling. One sees
that the expansion remains the same under the change $\Delta\to -\Delta$, so that such
pairs of curves connect smoothly through the critical point. (As pointed out above, each curve for a given $\Delta$ is
confined to one side of the critical point, such that $\nh/\nu_1$ and hence $\nh$ has the
same sign as $\Delta$, reflecting the constraint $\rt>0$.) Note that without pressure mixing also the
quadratic and third order terms in Eq.~\eqref{const_delta} are
$\Delta$-independent because $\tilde\tau_4$ and $\nu_2$ also vanish in this case, in addition to $\tilde\tau_2$. 

The cloud curve is obtained for
$\Delta=\pm1$ as
\bea
t &=& \frac{\tau_1}{\nu_1} \nh
- \tilde\tau_2'
\left(\frac{\nh}{\nu_1}\right)^2
- \tilde\tau_3\left|\frac{\nh}{\nu_1}\right|^{1/\bt}
\nn
& &{}-
\left[\tilde\tau_4+\tilde\tau_4'
- 2\left(\frac{\nu_2+\nu_2'}{\nu_1}\right)
\tilde\tau_2'\right]
\left(\frac{\nh}{\nu_1}\right)^3
\nn
&&{}- \tilde\tau_5
\left|\frac{\nh}{\nu_1}\right|^{1/\beta}.
\label{cloud}
\eea
and this result now applies for $\nh$ of arbitrary sign, i.e.\ for parent densities either side of $n_c$. One can check that the structure of the above expansion of the cloud curve is independent of pressure mixing: even if all pressure mixing mapping coefficients are set to zero, then generically none of the prefactors in Eq.\eqref{cloud} will vanish.
This is consistent with the result of Ref.\ \citenum{Belyakov2013}, where the authors developed a framework for multi-component fluids with fixed overall composition but without pressure mixing, obtaining essentially the same cloud curve structure as in Eq.~\eqref{cloud}, except for the third order term. This may have been omitted by accident in Ref.~\citenum{Belyakov2013} or dropped out because an intermediate expansion was truncated too early.

We note that in the case of mean-field ($\alpha=0$, $\beta=\bt=1/2$) rather than Ising criticality the second, third, and fifth terms in Eq.\ (\ref{cloud}) degenerate into a term proportional to $\nh^{2}$, and consequently the cloud curve becomes fully smooth around the critical
point, as expected. Otherwise, the term with the Fisher-renormalised exponent, $|\nh|^{1/\bt}$, is the first singular contribution. The latter may be challenging to detect in practice as it will be masked by the smooth variation given by the linear and the quadratic terms. One might then need to look at derivatives along the cloud curve, e.g.\ $d^3t/d\nh^3$ to see the singularity clearly as a
divergence, or at least $d^2t/d\nh^2$ to observe a cusp singularity.

In the special case of $\Delta=0$, one obtains the following generic form for the $50$--$50$ fractional volume line ($\nh$ must now have the same sign as $\nu_2$):
\bea
t &=& \frac{\tau_2}{\nu_2} \nh
- \left(\nu_3\frac{\tau_2}{\nu_2}-\tau_3\right)
\left(\frac{\nh}{\nu_2}\right)^{1/(2\bt)}
\nn
&&{}
- \left(\nu_5\frac{\tau_2}{\nu_2}-\tau_5\right)
\left(\frac{\nh}{\nu_2}\right)^{1/(2\beta)}
\label{ZeroDelta}
\eea
The slope $\tau_2/\nu_2$ of the linear piece in Eq.\ (\ref{ZeroDelta}) is equal to the slope $\tau_1/\nu_1$ of the linear piece in the cloud [Eq.\ (\ref{cloud})] because $\nu_2=-j_2\nu_1$ and $\tau_2=-j_2\tau_1$ as mentioned above. Note that one cannot interpolate smoothly to $\Delta=0$
once $\rt$ has been eliminated: the $\Delta\to 0$ limit
of Eq.~\eqref{const_delta} diverges. (It is important to note here that while all physical quantities vary smoothly with $\Delta$ {\em away} from the CP, this does not have to be the case in expansions around the CP.) The reason is that the expansion
in Eq.~\eqref{const_delta} is, effectively, in terms of $\nh/(\nu_1\Delta)$ and so is
valid in a range of width proportional to $\Delta$ that vanishes for
$\Delta\to 0$. 
Note the rather unexpected singularity exponent above: the singularity
for $\Delta=0$ is {\em stronger} than on the constant fractional volume lines  $\Delta\neq
0$, including the cloud curve.

The $\Delta=0$ constant fractional volume line is special also in that its shape depends sensitively on the presence or absence of pressure mixing. In the latter case one 
has $\nh=\nu_3\rt^{1/\bt}+\nu_5\rt^{1/\beta}$, which yields $\rt^{1/\bt} \sim \nh(1+\nh^{\bt/\beta-1}+\ldots)$ and hence (for $\nh/\nu_3>0$)
\be
t = \frac{\tau_3}{\nu_3} \nh-
\left(\nu_5\frac{\tau_3}{\nu_3}-\tau_5\right)\left(\frac{\nh}{\nu_3}\right)^{1/(1-\alpha)}
\label{ZeroDeltaWPM}
\ee
In this form where
$\rt$ has been eliminated the fact that pressure mixing changes
the leading singularity exponent from $1/(1-\alpha)$ to
$(1-\alpha)/(2\beta)$ may seem a little unexpected; however, it is quite natural
when looked at in terms of the vanishing of a number of contributions
in the $\rt$-expansions [Eqs.~\eqref{nh_exp} and \eqref{t_exp}]. Note that the slope of the linear term in Eq.~(\ref{ZeroDeltaWPM}) is generically {\em different} from the slope of the cloud curve at the critical point: without pressure mixing the $50$--$50$ fractional volume line departs from the critical point in a different direction, while in the presence of pressure mixing it starts off tangentially to the cloud curve.

To highlight the differences discussed above we now consider $\delta t_0\equiv t_{\rm cloud} - t_{\Delta=0}>0$, which is the temperature difference between the cloud curve and the $\Delta=0$ line at fixed parent density $\nh$. With pressure mixing this is the difference between Eq.~\eqref{cloud} and Eq.~\eqref{ZeroDelta}, which gives
\bea
\delta t_0 &=&
\left(\nu_3\frac{\tau_2}{\nu_2}-\tau_3\right)
\left(\frac{\nh}{\nu_2}\right)^{1/(2\bt)}
\nn
&&{}
+\left(\nu_5\frac{\tau_2}{\nu_2}-\tau_5\right)
\left(\frac{\nh}{\nu_2}\right)^{1/(2\beta)} + \ldots
\label{deltat0}
\eea
whereas in the case without pressure mixing it is the difference between Eq.~\eqref{cloud} (with modified prefactors due to pressure mixing being absent) and Eq.~\eqref{ZeroDeltaWPM}, i.e.
\bea
\delta t_0 &=&
\left(\frac{\tau_1}{\nu_1}-\frac{\tau_3}{\nu_3}\right) \nh
+
\left(\nu_5\frac{\tau_3}{\nu_3}-\tau_5\right)\left(\frac{\nh}{\nu_3}\right)^{1/(1-\alpha)}
\label{deltat0WPM}
\eea
With pressure mixing [Eq.~\eqref{deltat0}] there is \textit{no linear contribution}, reflecting the fact that the $50$--$50$ fractional volume line starts off tangential to the cloud curve, with the slopes of the linear pieces in Eqs.~\eqref{cloud} and \eqref{ZeroDelta} cancelling. Without pressure mixing this is no longer the case, leading to the linear piece in Eq.~\eqref{deltat0WPM}. This suggests that measurements of $\delta t_0$ as a function of $\nh$ could be useful probes of pressure mixing effects, in particular because the leading terms have exponents of $1/(2\bt)\simeq1.37$ and 1 respectively that are easy to distinguish. This is explored further in our numerical analysis in Section \ref{numerics}.

Looking next at the behaviour of the coexisting densities, these are
easy to determine since Eq.~\eqref{nh_pm} has, for constant $\Delta$, the same
structure as the expansion of the parent density, Eq.~\eqref{nh_exp}. Therefore one can read off directly the expression for $t$ vs.\ $\nh_\pm$, but as this is rather long we defer it to Appendix \ref{MoreCoexDens}. One useful special case is $t$ vs.\ $\nh_\pm$ for $\Delta=\mp1$, which gives  the shadow curve. As can be seen in Appendix \ref{MoreCoexDens}, it has the same structure as the cloud curve [Eq.~\eqref{cloud}], but with
different coefficients (obtained via $\nu_1\to \nu_1-2\ltt$, $\nu_2\to
\nu_2-g_2$, $\nu_2'\to \nu_2'-g_2$, $\nu_4\to \nu_4-g_4$, $\nu_4'\to
\nu_4'-g_4'$).

For the other interesting special case of $\Delta=0$ one obtains the
`$50$--$50$ coexistence curves', i.e.\ the temperature dependence of the coexisting densities obtained from parents on the $50$--$50$ fractional volume line:
\bea
t &=& \tau_2
\left(\frac{\nh_\pm}{\ltt}\right)^2
+\tau_3 \left|\frac{\nh_\pm}{\ltt}\right|^{1/\bt}
+2\frac{\nu_2\tau_2}{\ltt}\left(\frac{\nh_\pm}{\ltt}\right)^3
\nn
&&{}
+\tau_5 \left|\frac{\nh_\pm}{\ltt}\right|^{1/\beta}.
\label{coex_Delta0}
\eea
The two curves $t(\nhp)$ and $t(\nhm)$ are symmetric in the first two
leading terms, but then the asymmetry appears. Without pressure mixing one has $\tau_2=0$ and so the leading quadratic as well as the cubic terms are both absent. 

It is useful to convert the above results into expressions of densities versus temperature to connect with our discussion of the diameter in the introduction. The $\Delta=0$ line is the equal volume diameter and, by inverting \eqref{ZeroDelta} and \eqref{ZeroDeltaWPM}, respectively, is given by
\bea
\nh &=&  \frac{\nu_2}{\tau_2}t
+ \left(\nu_3 - \tau_3\frac{\nu_2}{\tau_2}\right)
\left(\frac{t}{\tau_2}\right)^{1/(2\bt)}
\nn
&&{}
+ \left(\nu_5 - \tau_5\frac{\nu_2}{\tau_2}\right)
\left(\frac{t}{\tau_2}\right)^{1/(2\beta)}.
\label{nh_vs_t_Delta0}
\eea
with pressure mixing and by
\be
\nh =  \frac{\nu_3}{\tau_3}t
- \left(\tau_5\frac{\nu_3}{\tau_3}-\nu_5\right)
\left(\frac{t}{\tau_3}\right)^{1/(1-\alpha)}
\label{nh_vs_t_Delta0WPM}
\ee
without. Note that in both cases the leading term is linear, in contrast to the situation in the monodisperse case discussed in the introduction, and it is only the exponent of the first subleading term that signals the presence or absence of pressure mixing.

Generally, if we consider $\Delta$ being varied from $1$ to $-1$, the curve $t(\nh)$
will be deformed from the high-density to the low-density part of the cloud
curve (see Fig.\ \ref{CloudShadowPlot}); similarly $t(\nhp)$ interpolates between the high-density parts
of the cloud curve and of the shadow curve, while $t(\nhm)$
interpolates between the low-density parts of the shadow curve and of
the cloud curve.

\section{The monodisperse case}
\label{mono}

We have seen above that the complete scaling predictions for the equal volume diameter are different from those reviewed in the introduction for monodisperse systems, whether with or without pressure mixing. Ostensibly, however, our analysis is valid for an arbitrary number of mixture components $M$. What, then, is different about the monodisperse case ($M=1$)?

One answer is that when there is only a single chemical potential, the conditions in Eqs.~\eqref{cond0_mix}--\eqref{cond2_mix} are
sufficient to determine $\pt$, $t$, $\mu$ and these must therefore be
smooth functions of $\tt=-a\rt^{1/\beta}$. (A similar comment can be found in Ref.\ \citenum{ISI:000345509900008}.) In our $\rt$-expansions, Eqs.~\eqref{ph_exp}--\eqref{muh_exp}, this means that the only nonzero coefficients are $\pi_5$, $\tau_5$, and $\mv_5$, the latter being a scalar for $M=1$. (One can easily
check this from the explicit conditions for the coefficients; e.g.\ the first order conditions in Eqs.~\eqref{firstorderconditions1}--\eqref{firstorderconditions4} have the obvious solution
$\pi_1=\tau_1=\mvs_1=0$, $\nu_1=\ltt$ in the monodisperse case.) Since
$\rt\sim|\tt|^{\beta}\sim|t|^{\beta}$, the coexisting densities then have the
standard expansion $\nh_{\pm}\sim \pm \rt+\rt^2 +\rt^{1/\bt}+\rt^{1/\beta} \sim
\pm|t|^{\beta}+|t|^{2\beta}+|t|^{1-\alpha}+|t|$ and the first term
cancels from the diameter.

In the polydisperse case the solution above does not work since e.g.\
in the first order conditions in Eqs.~\eqref{firstorderconditions1}--\eqref{firstorderconditions4} the vectors $\f$ and $\lt_2$ will
generically not be parallel. Alternatively, one can go back to the
dilution line constraint in Eq.~\eqref{coex_cond}: there are singular (in $\tt$)
terms on the right-hand side, proportional to $\rt$, $\rt^2$ and $\st$. If
$\muh$ and $t$ depended smoothly on $\tt$, these singular terms would
always dominate and so push the system off the required dilution line.
To avoid this, $\muh$ and $t$ themselves need to contain terms proportional to
$\rt$, $\rt^2$ and $\st$. In the monodisperse case there
is no dilution line constraint and so no such requirement.

One might argue that the above discussion cannot be taken literally for most colloidal systems, because of the difficulty of producing the truly identical particles that a description in terms of a single species ($M=1$) in principle requires. A more realistic endeavour would be to make the distribution of particle sizes (say) narrower and narrower by improving experimental protocols. In the limit, one would then still expect to retrieve monodisperse phase behaviour, but this would result from an $M$-species system where the differences between species have become very small. How is this second route to the monodisperse limit achieved in our approach?

Taking the true monodisperse limit where one has $M$ species of particles that are physically identical, one can think of the different species as being identified by different colours but with this colour having no effect on the physical behaviour.
We show in Appendix \ref{equivalence} that in this description of a physically monodisperse -- but `colour polydisperse' -- system, the vectors $\f$ and $\lt_2$ will always be parallel. As discussed above, there is then no need for the additional scaling terms required in a system that is physically polydisperse, and standard monodisperse scaling behaviour follows. (We note as an aside that similar conceptual issues, related to the physical relevance of polydispersity in the limit when particle species become very similar, arise in determining the configurational entropy of glasses, see e.g.\ Ref.\ \citenum{BerthierEntropyPoly}.)

The derivation in Appendix \ref{equivalence} is effectively a method by which one can map \textit{the mixing coefficients} of a monodisperse fluid onto a new set of mixing coefficients for a colour polydisperse fluid with nominally $M>1$ components, in such a way that both systems are physically equivalent with regards to their critical behaviour. The method is in fact more general, allowing one to map the mixing coefficients of a fluid with $M'$ components onto a physically identical fluid with nominally $M>M'$ components. 

\section{Coexistence curves}
\label{coexistence}

Next we look at conventional coexistence curves, which are obtained from a parent of fixed density $\nh$ by varying temperature. With $\nh$ fixed, $\Delta$ then has to vary appropriately to maintain particle conservation.

We start by separating
$\Delta\rt$-terms and $\rt$-terms in Eq.~\eqref{nh_exp} for $\nh$:
\bea
\nh - \nu_2\rt^2 - \nu_3 \rt^{1/\bt}  - \nu_5 \rt^{1/\beta} &=&
\label{parentvsrt}
\nu_1\Delta\rt + \nu_2'(\Delta\rt)^2  
\\&&{}
+ \nu_4(\Delta\rt)\rt^2+\nu_4'(\Delta\rt)^3
\nonumber
\eea
This can be solved perturbatively for $\Delta\rt$:
\bea
\Delta\rt &=&
\frac{\nh}{\nu_1}
-\frac{\nu_2}{\nu_1}\rt^2-\frac{\nu_2'}{\nu_1}\left(\frac{\nh}{\nu_1}\right)^2
-\frac{\nu_3}{\nu_1}\rt^{1/\bt}
\nn
&&{}
-\left(\frac{\nu_4}{\nu_1}-2\frac{\nu_2}{\nu_1}\frac{\nu_2'}{\nu_1}\right)
\rt^2\frac{\nh}{\nu_1}
\nn
&&{}
-\left[\frac{\nu_4'}{\nu_1}-2\left(\frac{\nu_2'}{\nu_1}\right)^2\right]
\left(\frac{\nh}{\nu_1}\right)^3
-\frac{\nu_5}{\nu_1}\rt^{1/\beta}
\label{Delta_rt_soln}
\eea
We have used that since $\nh=\nu_1\Delta\rt$ to leading order, $\nh$ is
never larger than $\sim\rt$, so that we are sure to have all relevant
terms if we treat $\nh$ as proportional to $\rt$ and then expand.
Now one inserts (\ref{Delta_rt_soln}) into the expansion for $t$ in Eq.~\eqref{t_exp}:
\bea
t &=& \frac{\tau_1}{\nu_1} \nh 
-\tilde\tau_2'\left(\frac{\nh}{\nu_1}\right)^2 
-\tilde\tau_3\rt^{1/\bt}
\label{tvsnhandrt}
\\
&&{}
-\left(\tilde\tau_4-2\frac{\nu_2}{\nu_1}\tilde\tau_2'\right)
\frac{\nh}{\nu_1}\rt^2
-\tilde\tau_4'
\left(\frac{\nh}{\nu_1}\right)^3
-\tilde\tau_5\rt^{1/\beta}
\nonumber
\eea
The leading linear and quadratic terms can be cancelled by switching from $t$ to the  temperature difference from
the cloud point temperature, $\delta t \equiv t_{\rm cloud}-t$. Here $t_{\rm cloud}$ is given by Eq.~\eqref{cloud} and is fixed by the given parent density $\nh$. In terms of $\delta t$, Eq.~\eqref{tvsnhandrt} takes the simpler form
\bea
\delta t &=& 
\tilde\tau_3
\left[\rt^{1/\bt}- \left|\frac{\nh}{\nu_1}\right|^{1/\bt}\right]
\nn
& &{}
+\left(\tilde\tau_4-2\frac{\nu_2}{\nu_1}\tilde\tau_2'\right)
\frac{\nh}{\nu_1}\left[\rt^2 - \left(\frac{\nh}{\nu_1}\right)^2\right]
\nn
& &{}
+ \tilde\tau_5
\left[\rt^{1/\beta}- \left|\frac{\nh}{\nu_1}\right|^{1/\beta}\right].
\label{deltat}
\eea
(Note that $\delta t$ is positive by definition and so needs
to increase with $\rt$; thus $\tilde\tau_3$ should be
positive.) 

Our goal is to find the coexisting densities $\nh_\pm$. For these one uses Eq.~\eqref{nh_pm2} with Eq.~\eqref{Delta_rt_soln} inserted, to obtain $\nh_\pm$ as a function of $\rt$ for fixed $\nh$. Now $\rt$ needs to be eliminated between the resulting expression [see (\ref{nhpm_coex}) in Appendix \ref{MoreCoexDens}] and Eq.~(\ref{deltat}). As $\rt$ only appears at second and higher order in Eq.~(\ref{deltat}), it suffices to find it from (\ref{nhpm_coex}) to linear order in terms of $\nh\onetwo$, resulting in the simple expression
\be
\rt = \pm
\left(\frac{\nh\onetwo}{\ltt}+\x\nh\right)
\label{rt_vs_nhonetwo}
\ee
where $\x\equiv1/\nu_1-1/\ltt$. Inserting into Eq.~\eqref{deltat} produces finally
\bea
\delta t &=& 
\tilde\tau_3
\left(\left|\frac{\nh\onetwo}{\ltt}+\x\nh\right|^{1/\bt}- \left|\frac{\nh}{\nu_1}\right|^{1/\bt}\right)
\nn
& &{}+\left(\tilde\tau_4-2\frac{\nu_2}{\nu_1}\tilde\tau_2'\right)\frac{\nh}{\nu_1}
\left[\left(\frac{\nh\onetwo}{\ltt}+\x\nh\right)^{2}-
\left(\frac{\nh}{\nu_1}\right)^{2}\right]
\nn
& &{}+ \tilde\tau_5
\left(\left|\frac{\nh\onetwo}{\ltt}+\x\nh\right|^{1/\beta}-
\left|\frac{\nh}{\nu_1}\right|^{1/\beta}\right).
\label{CoexCurveFixedDens}
\eea

One sees that at the onset of phase coexistence ($\delta t=0$), $\nh\onetwo/\ltt+\x\nh$
must equal $\nh/\nu_1$ or $-\nh/\nu_1$ to leading order. This gives $\nh\one=\nh$ and
$\nh\two=\nh(1-2\ltt/\nu_1)$ or vice versa; the latter prefactor is
consistent with the ratio of the slopes of cloud and shadow curves at
the critical point, which can be read off from Eqs.\ \eqref{cloud} and \eqref{shadow}. 

The above expressions simplify considerably for the case $\nh=0$. Eq.~\eqref{CoexCurveFixedDens} then gives the critical coexistence curve as
\bea
|t| = \delta t &=& 
\tilde\tau_3 \left|\frac{\nh\onetwo}{\ltt}\right|^{1/\bt}
+ \tilde\tau_5 \left|\frac{\nh\onetwo}{\ltt}\right|^{1/\beta}
\nn
\label{CoexCurveCritical}
\eea
The leading term shows the expected Fisher-renormalised order parameter exponent $\bt$. More interestingly, the structure of this conventional coexistence curve has no obvious signatures of pressure mixing, a situation rather different from the $50$--$50$ ($\Delta=0$) coexistence curve in Eq.~\eqref{coex_Delta0}.

We next consider the temperature variation of the fractional volumes of the coexisting phases, more specifically their difference $\Delta$. In Eqs.~\eqref{Delta_rt_soln} and \eqref{deltat} we already have
$\Delta\rt$ and $\delta t$, both as functions of $\rt$ (for fixed
$\nh$). Solving the second of these equations for $\rt$ gives to leading order
\be
\rt^{1/\bt} = \frac{\delta t}{\tilde\tau_3} +
\left|\frac{\nh}{\nu_1}\right|^{1/\bt}
\label{rt2_vs_dt}
\ee
If we then keep only the
leading terms in $\Delta\rt$ and $\rt$ and write the ratio between them we find
\be
\Delta = 
\frac{\nh-\nu_2(\delta t/\tilde\tau_3)^{2\bt}}
{\nu_1\left[\delta t/\tilde\tau_3 +
	\left|\nh/\nu_1\right|^{1/\bt}\right]^{\bt}}
\label{deltavst}
\ee
\\
For off-critical parents ($\nh\neq 0$) this starts off at $\pm 1$ as it should at the cloud point and then
decreases (in modulus), scaling for $\nh^{1/\bt}\ll\delta t \ll \nh^{1/(2\bt)}$ as
$\nh(\delta t)^{-\bt}$. The behaviour changes when $\delta t\sim \nh^{1/(2\bt)}$
and crosses over to $\Delta \approx-(\nu_2/\nu_1)(\delta
t/\tilde\tau_3)^{\bt}$ for $\delta t\gg \nh^{1/(2\bt)}$. Only the latter regime
is present for the critical coexistence curve ($\nh=0$). For $\nh\neq
0$, the crossover between the different regimes implies that on one
side of the critical point $\Delta$ will always depend
non-monotonically on $\delta t$. (To find which side will have that behaviour we have to look at the sign of
$-(\nu_2/\nu_1)$: if this is positive, $\Delta$ is increasing for
large $\delta t$ and so the non-monotonicity occurs on the
high-density side of the critical point, where $\Delta$ initially \textit{decreases} from $1$.)

Without pressure mixing, where $\nu_2$ vanishes, the corresponding expression is
\be
\Delta = 
\frac{\nh-\nu_3(\delta t/\tilde\tau_3)}
{\nu_1\left[\delta t/\tilde\tau_3 +
	\left|\nh/\nu_1\right|^{1/\bt}\right]^{\bt}}
\label{deltavstWPM}
\ee
which scales as $\nh(\delta t)^{-\bt}$ for $\nh^{1/\bt}\ll\delta t\ll \nh$,
changing when $\delta t\sim |\nh|^{1/\bt}$ and crossing over to $(\delta t)^{1-\beta}$ for $\delta t\gg\nh$; similar comments
about non-monotonicity apply as above, but now one has to look at the sign of $-(\nu_3/\nu_1)$.

Finally we ask about the behaviour of the midpoint diameter of the coexistence
curves, as well as about its analogue at off-critical parent densities. For any fixed $\nh$, we define $\nbar\equiv\half(\nh\one+\nh\two)$ as in the introduction. The steps sketched in Appendix \ref{MoreCoexDens} lead to
\be
\nbar = \frac{\nu_2\ltt}{\nu_1}\left(\frac{\delta t}{\tilde\tau_3}\right)^{2\bt}-2\bt\frac{\nu_2\ltt}{\nu_1}\frac{\tilde\tau_5}{\tilde\tau_3}\left(\frac{\delta t}{\tilde\tau_3}\right)^{2\bt+\alpha/(1-\alpha)}
\label{diameter}
\ee
In the case without pressure mixing, we have
\be
\nbar=\frac{\ltt\nu_3}{\tilde\tau_3\nu_1}\delta t + \frac{\ltt}{\nu_1}\left(\nu_5-\nu_3\frac{\tilde\tau_5}{\tilde\tau_3}\right)\left(\frac{\delta t}{\tilde{\tau_3}}\right)^{1/(1-\alpha)}
\label{diameterWPM}
\ee
Compared to the results quoted in the introduction, one sees that the leading exponents are Fisher-renormalised because of the presence of polydispersity, giving ${2\beta/(1-\alpha)=2\bt}$ with pressure mixing and ${(1-\alpha)/(1-\alpha)=1}$ without.

A comparison with the equal volume diameter results is also instructive: Eq.~\eqref{nh_vs_t_Delta0WPM} shows that without pressure mixing both diameters have the same scaling behaviour but with different prefactors. With pressure mixing, only the midpoint diameter has a leading singular term, while the equal volume diameter starts off linearly as shown by Eq.~\eqref{ZeroDeltaWPM}. It is worth emphasizing also that the midpoint diameter, which is obtained by cooling a parent with the critical density, always relates to temperatures below the critical point, $t<0$. The equal volume diameter has no such restriction -- it only has to lie within the coexistence region, i.e.\ below the cloud curve -- as we will see in the numerical illustrations in Sec.~\ref{numerics}.

Above we have written the midpoint diameter density as a function of temperature, to ease comparison with treatments elsewhere in the literature. For consistency with the way we have expressed our results for other phase diagram loci we give the corresponding inverted relations here, which read
\be
\delta t = \tilde\tau_3\left(\frac{\nu_1\nbar}{\nu_2\ltt}\right)^{1/(2\bt)}+\tilde\tau_5\left(\frac{\nu_1\nbar}{\nu_2\ltt}\right)^{1/(2\beta)}
\label{diameterv2}
\ee
with pressure mixing and
\bea
\delta t = \tilde\tau_3\left(\frac{\nu_1\nbar}{\nu_3\ltt}\right)+\left(\tilde\tau_5-\tilde\tau_3\frac{\nu_5}{\nu_3}\right)\left(\frac{\nu_1\nbar}{\nu_3\ltt}\right)^{1/\left(1-\alpha\right)}
\label{diameterWPMv2}
\eea
in the case without. As written, the results in Eqs.~\eqref{diameter}, \eqref{diameterWPM}, \eqref{diameterv2}, and \eqref{diameterWPMv2} are valid only for $\nh=0$, but their off-critical versions can be obtained simply via the replacement $\nbar\rightarrow\nbar-\nh\left(1-\frac{\ltt}{\nu_1}\right)$.

\section{Numerics}
\label{numerics}
 
In this section we compare the scaling expansions derived above to direct numerical evaluation of the complete scaling theory. 
For a given set of numerical values for the mixing coefficients one needs to solve the  mapping equations (\ref{pt_mix})--(\ref{ht_mix}) and the dilution line (or particle conservation) constraint (\ref{dilution}). The relation between the Ising variables is given by the scaling relation (\ref{pt_scaling_form}) together with the coexistence-region condition $\htt=0$. 
In solving this full system we will not make any further approximations; in particular we will retain $\pt$ rather than neglecting it as subleading, and we will use the general expressions for $\bm \rh$ from Eq.\ (\ref{rh_gen}) without further expansion. (By setting $\tilde{h}=0$ we are neglecting the asymmetry terms from Eq.~\eqref{ht_scaling}; these are subleading even compared to the effect of allowing $\tilde{p}$ to be nonzero.) For the required explicit form of the scaling relation (\ref{pt_scaling_form}) we have used the `parametric linear model' of Ref.~\citenum{schofield1969correlation}.

While in the analytical development of the previous sections it was convenient to use $\rt$ to parameterise the various curves in the phase diagram, for the numerics we prefer $\tt$ as this enters directly in the scaling form (\ref{pt_scaling_form}).
One ends up with a system of $M+3$ equations (mapping equations plus dilution line constraint) and $M+4$ variables, the extra variable in addition to the $M+3$ variables $\nh, \ph, t$, and $\muh$ being precisely $\tt$. For each value of this variable we solve the full system of equations to generate a point in the space of the physical fluid variables. With this approach we can check both the exponents \textit{and prefactors} of the analytical expansions obtained above. In particular we will discuss the effects of pressure mixing. We will see that these generally become weaker as one moves away from the critical point, leading to crossovers to behaviour characteristic of a system without pressure mixing.

\begin{figure}
\hspace{-0.5cm}
		\includegraphics[clip,width=\columnwidth]{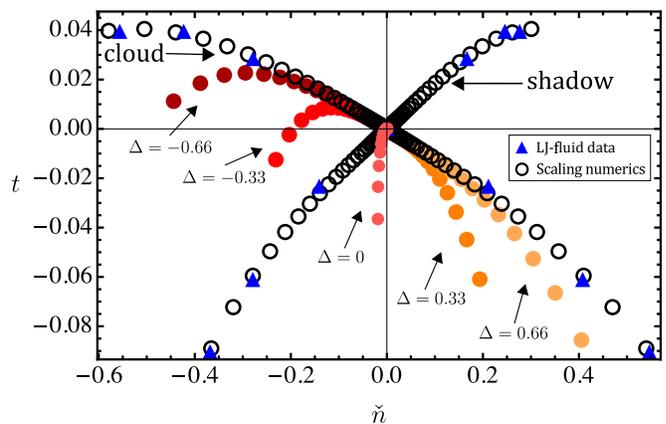}
		\caption{Cloud and shadow data from polydisperse Lennard-Jones (LJ) fluid simulations in Ref.\ \citenum{LJNigelPeter} (triangles) and from our numerical solution (empty circles). From warmer to cooler colours, we also show data for five constant fractional volume lines, for values of $\Delta$ as shown. (In the convention used in later figures, the strength of pressure mixing here is $f_{\rm pm}=1$.)
}
\label{CloudShadowPlot}
\end{figure}

We started by finding fitted values for the mixing coefficients by comparison with cloud and shadow data obtained in Ref.\ \citenum{LJNigelPeter} from computer simulations of a Lennard-Jones (LJ) fluid with `amplitude' polydispersity. The results as well as the details of the fitting method are described in Appendix \ref{AppNum}. With these fitted mixing coefficients the numerical solution led to the cloud and shadow data in Fig.\ \ref{CloudShadowPlot}; notice the good agreement with the LJ fluid data. We omitted in the fit and in the comparison in Fig.\ \ref{CloudShadowPlot} those data points from Ref.\ \citenum{LJNigelPeter} that were too far from the CP to allow a meaningful comparison with a scaling theory for the critical behaviour.
From warmer to cooler colours, we also show data for the numerically calculated constant fractional volume lines ($t$ vs.\ $\nh$) for $\Delta=-0.66$, $-0.33$, $0$, $0.33$ and $0.66$, respectively.
Notice that the $\Delta=0$ curve, which is the equal volume diameter, initially moves left and up from the CP, towards higher temperatures and lower densities. It shares this behaviour with the constant fractional volume lines for negative $\Delta$ as discussed in detail below.

For further numerical evaluation it is useful to be able to vary the strength of pressure mixing.
We do this by scaling the values of the pressure mixing coefficients according to
$j_1 \rightarrow f_{\text{pm}}\, j_1$ and $j_2 \rightarrow f_{\text{pm}} \,j_2$, where the factor $f_{\text{pm}}$ is then a `pressure mixing fraction'. (With this definition, $f_{\rm pm}=1$ in Fig.\ \ref{CloudShadowPlot}.) Changing $f_{\text{pm}}$ in the range $0$ to $4$ we obtain data that are physically reasonable and only slightly off the LJ data from Ref.~\citenum{LJNigelPeter}. But predictions change close to the critical point, as we will explore shortly.

\begin{figure}[h!]
	\includegraphics[width=\columnwidth]{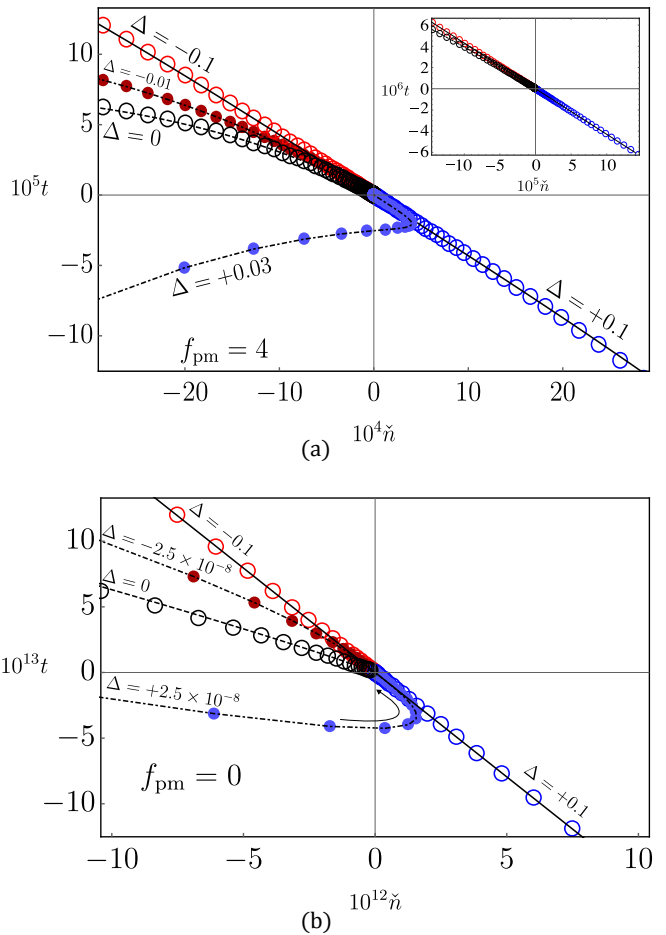}
	\caption{Constant fractional volume lines. Linear scale plot of numerical data (circles) and theoretical predictions (lines) for the lines defined by $\Delta=0$ and ${\Delta=\pm0.1}$.
(a) Strong pressure mixing, $f_{\rm pm}=4$. Inset: zoom on the region close to the CP. 
(b) No pressure mixing, $f_{\rm pm}=0$; notice the distinct slope of the equal volume diameter ($\Delta=0$ line). Additional numerical data (small filled symbols and dash-dotted lines) are shown for two small $|\Delta|$ in both (a) and (b), with $\Delta$-values chosen in order to illustrate the smooth $\Delta$-dependence of the curves away from the CP.}
	\label{Linear}
\end{figure}

We have selected a number of key features of the scaling formulae obtained above for which we will present supporting numerical data. Firstly, Fig.\ \ref{Linear} illustrates how the behaviour of the constant fractional volume lines ($t$ vs.\ $\nh$ for fixed $\Delta=0, \pm0.1$) 
depends on pressure mixing. 
For strong pressure mixing (Fig.\ \ref{Linear}a) we can clearly see that, although the curve for $\Delta=0$ separates from the one with $\Delta=-0.1$ as it moves away from the CP (main graph), the two curves become tangential to each other at the CP, i.e.\ they have the same slope there (inset). 
This agrees with the prediction for the prefactors of the leading linear terms in Eqs.\ \eqref{const_delta} and \eqref{ZeroDelta}. The curve with $\Delta=0.1$ has the same initial slope but departs from the CP in the opposite direction, towards lower temperatures and larger parent densities. Two additional lines in Fig.\ \ref{Linear}a show numerical results for small $|\Delta|$. These clarify that, as noted before, all properties are smooth in $\Delta$ {\em away} from the CP, but cross over to the discontinuous change at $\Delta=0$ in the direction of departure from the CP itself .

Fig.\ \ref{Linear}b contrasts these observations with the case without pressure mixing, obtained by setting $f_{\rm pm}=0$. The curves for $\Delta=\pm 0.1$ again show the same slope at the CP and depart in opposite directions. However, the $\Delta=0$ line now has a different slope at the CP and separates linearly from the other lines, consistent with the different prefactors of the linear terms in Eqs.\ \eqref{ZeroDelta} and \eqref{ZeroDeltaWPM} and the discussion in Sec.~\ref{fractional} above. Two further small $|\Delta|$ lines again illustrate how these discontinuous changes at the CP connect to the smooth $\Delta$-dependences away from the CP.

\begin{figure}[h!]
	\includegraphics[width=1\columnwidth]{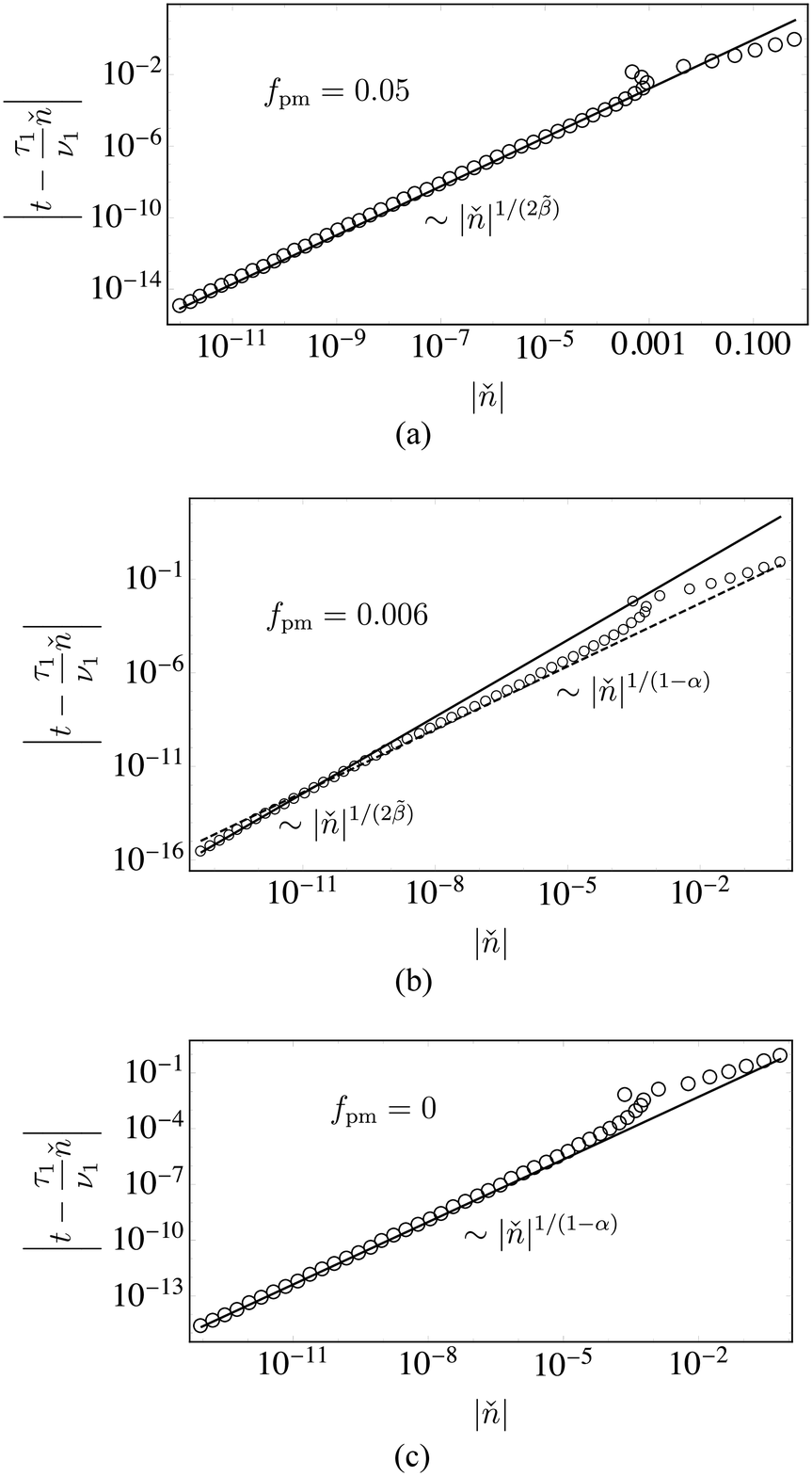}
	\caption{Equal volume diameter ($\Delta=0$). Log-log plots of numerical data for $|t-\frac{\tau_1}{\nu_1}\nh|$ vs.\ $|\nh|$, i.e.\ with predicted leading term removed (empty circles). 
		All plots show a change of sign in $\nh$ at higher $|\nh|$. Lines are theoretical predictions. (a) Moderate pressure mixing, $f_{\text{pm}}=0.05$, shows the leading power law term with exponent $1/(2\bt)\simeq1.37$. (b) Weak pressure mixing, $f_{\text{pm}}=0.006$, produces a crossover between power laws with pressure mixing (solid line) and without (dashed line, exponent $1/\left(1-\alpha\right)\simeq1.12$).
		(c) No pressure mixing, $f_{\text{pm}}=0$.
	}
	\label{ZeroDeltaFigure}
\end{figure}

Fig.\ \ref{ZeroDeltaFigure} illustrates the nonlinear terms in the equal volume diameter ($\Delta=0$ constant fractional volume curve), by showing this diameter on log scales with the leading linear $\nh$-dependence predicted from Eq.~\eqref{ZeroDelta} taken off.
For high values of $f_{\rm pm}$ we therefore see a leading nonlinear term with exponent $1/(2\bt)$, in Fig.\ \ref{ZeroDeltaFigure}a. Note that for $f_{\text{pm}}=1$ the sub-sub-leading term in Eq.\ (\ref{ZeroDelta}), with exponent $1/(2\beta)$, would kick in at a value of $|\nh|$ well inside the range shown. In order to see the first sub-leading term with exponent $1/(2\bt)$ clearly we therefore used 
$f_{\text{pm}}=0.05$. A small residue of this competition remains, causing the slight upward shift of the numerical data compared to the theory at low $|\nh|$. 

As we decrease $f_{\rm pm}$ (Fig.\ \ref{ZeroDeltaFigure}b), a crossover to the result without pressure mixing, Eq.~\eqref{ZeroDeltaWPM}, becomes visible inside our $|\nh|$-range. However, this occurs where the dominant term in  Eq.~\eqref{ZeroDeltaWPM} is already the one with exponent $1/(1-\alpha)$. At $f_{\rm pm}=0$, finally, pressure mixing effects disappear and we observe only the term with exponent $1/(1-\alpha)$ (see Fig.\ \ref{ZeroDeltaFigure}c). As in Fig.\ \ref{ZeroDeltaFigure}b, there is in principle a linear contribution  that arises because we are subtracting the linear prediction {\em with} pressure mixing, which has a different prefactor and so cannot cancel the linear term {\em without} pressure mixing. This should dominate at small $|\nh|$ but is quantitatively too small to be visible. 
We emphasize that in all cases the numerical data agree with the theoretical predictions agree not just in exponent (slope) but also in prefactor. 

\begin{figure}
	\includegraphics[clip,width=1\columnwidth]{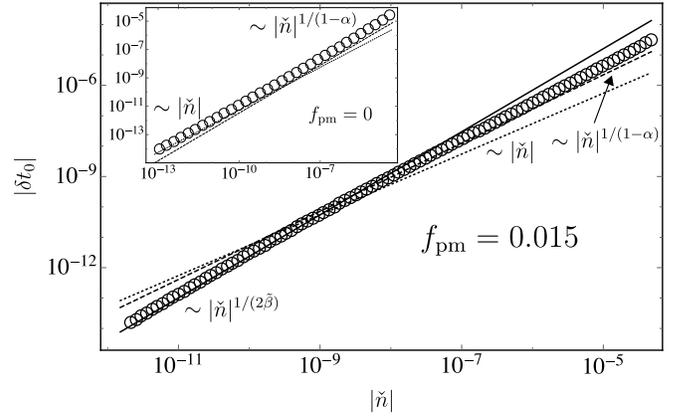}%
	\caption{Temperature difference between cloud curve and equal volume diameter. Log-log plots of numerical data for $|\delta t_0|$ vs.\ $|\nh|$ at $f_{\rm pm}=0.015$ (empty circles). Solid line: contribution with exponent $1/(2\bt)\simeq1.37$ from the theoretical prediction. A crossover to the regime without pressure mixing is seen, where terms with exponents $1$ (dotted) and $1/(1-\alpha)\simeq1.12$ (dashed) compete, kick in around the same value of $|\nh|$. 
Inset: Without pressure mixing, $f_{\rm pm}=0$, the term with exponent $1/(2\bt)$ is absent.
}
	\label{CrossOverDifference}
\end{figure}

As discussed above in Section \ref{fractional},  Eq.~\eqref{deltat0} shows that in the `with pressure mixing' regime the temperature difference $\delta t_0=t_{\rm cloud}- t_{\Delta=0}$ between the cloud curve and the equal volume diameter should have no linear term in $\nh$; the leading contribution is singular, with exponent $1/(2\bt)$ from Eq.~\eqref{deltat0}. This is exactly what can be observed in Fig.\ \ref{CrossOverDifference}. Subsequently, we would expect to see the term with exponent $1/(2\beta)$ as one moves away from CP. However, \textit{before} this happens,
a crossover to the `without pressure mixing' regime occurs, where 
there is a leading linear piece from Eq.~\eqref{deltat0WPM}. This is however largely masked by the next term with exponent $1/(1-\alpha)$.
If pressure mixing is switched off completely ($f_{\rm pm}=0$), the linear contribution is clearly visible (see inset of Fig.\ \ref{CrossOverDifference}) as well as the crossover to $|\nh|^{1/(1-\alpha)}$.

\begin{figure}
	\includegraphics[width=1\columnwidth]{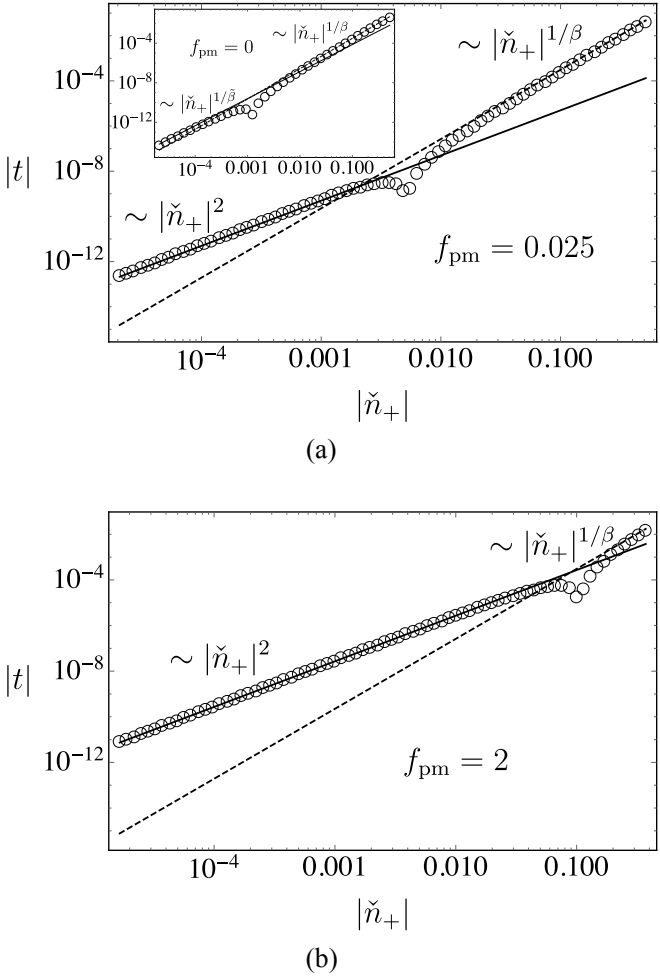}
	\caption{Equal fractional volume coexistence curve. (a) Log-log plot of numerical data for $|t|$ vs.\ $|\nh_{+}|$ for $\Delta=0$ at $f_{\text{pm}}=0.025$ (empty circles), showing a change of sign in $t$. Solid and dashed lines: theoretical predictions for the terms with exponent $2$ and $1/\beta\simeq3.07$, respectively. Inset: No pressure mixing, $f_{\text{pm}}=0$. As predicted, only contributions with exponents $1/\bt\simeq 2.73$ (solid) and $1/\beta$ (dashed) can be observed. (b) Similar to (a) but now with stronger pressure mixing, $f_{\text{pm}}=2$. The characteristic quadratic term is now visible as far as $\nh\simeq10\%$ from the CP.
}
	\label{CoexCurveZeroDeltaAllfpm}
\end{figure}

For our last set of numerical results for constant fractional volume we look in Fig.\ \ref{CoexCurveZeroDeltaAllfpm} at the $\Delta=0$ coexistence curves of $t$ vs.\ $\nh_{\pm}$,
in order to verify Eq.~\eqref{coex_Delta0}. 
Even with weak pressure mixing ($f_{\text{pm}}=0.025$) the leading quadratic term can easily be discerned, 
as can a contribution that survives even without pressure mixing, with exponent $1/\beta$. 
The terms with intermediate exponents $3$ and $1/\bt$ cannot be seen as they are quantitatively too small. This is also why we do not plot the $\nh_-$ branch, which would differ from $\nh_+$ only by the small third order term.
The inset shows the case $f_{\text{pm}}=0$: now only the terms with exponents $1/\bt$ and $1/\beta$ can be observed. This is as expected from Eq.~\eqref{coex_Delta0} since without pressure mixing  the terms with exponents $2$ and $3$ drop out. Fig.\ \ref{CoexCurveZeroDeltaAllfpm}b is similar to Fig.\ \ref{CoexCurveZeroDeltaAllfpm}a, but now for larger $f_{\text{pm}}=2$. The quadratic term that signals pressure mixing is clearly visible starting from $\nh\simeq10\%$ and extending for several decades towards the CP, suggesting that it could be amenable to relatively straightforward experimental verification.

\begin{figure}
	\includegraphics[width=\columnwidth]{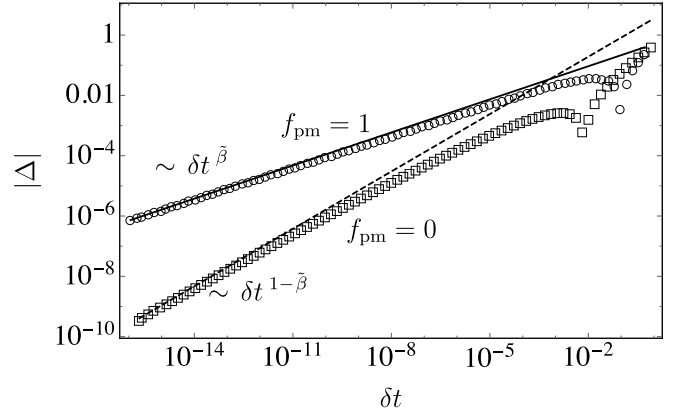}%
	\caption{Difference in fractional volumes of coexisting phases produced from critical parent ($\nh=0$). Log-log plots of numerical data for $|\Delta|$ vs.\ $|\delta t|$; $\Delta$ changes sign at larger 
$\delta t$. Data with pressure mixing ($f_{\text{pm}}=1$, empty circles) and without 
($f_{\text{pm}}=0$, empty squares) agree well with the theoretically predicted power laws with exponents
$\bt\simeq0.37$ and $1-\bt\simeq0.63$, respectively.
}
	\label{CriticalDeltaVSTemp}
\end{figure}

Now we move on to the numerical results for fixed parent density $\nh$. At $\nh=0$ we first consider (see Fig.\ \ref{CriticalDeltaVSTemp}) the dependence of the fractional volume parameter $\Delta$ on the temperature difference to the CP, $\delta t=-t$.
Data for 
$f_{\text{pm}}=1$ and $f_{\text{pm}}=0$ 
clearly show the 
exponents and prefactors for the cases with and without pressure mixing as predicted by Eqs.~\eqref{deltavst} and \eqref{deltavstWPM}, respectively. Note that close to the CP the difference $|\Delta|$ between the fractional volumes of the two coexisting phases is orders of magnitude larger with pressure mixing than without, suggesting a potential route for experimental detection of pressure mixing effects that would not require precise exponent measurements.

\begin{figure}
	\includegraphics[width=\columnwidth]{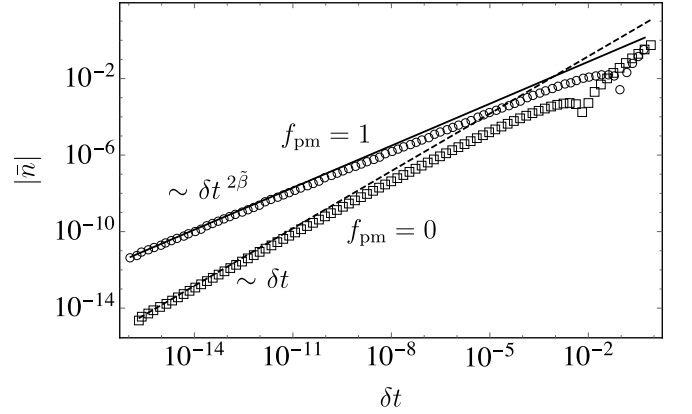}%
	\caption{Midpoint diameter. Log-log plots of numerical data for $|\bar{n}|$ vs.\ $|\delta t|$ for fixed $\nh=0$. 
For $f_{\text{pm}}=1$ (pressure mixing, empty circles), the solid line shows our prediction for the power law, with exponent $2\bt\simeq0.73$. For $f_{\text{pm}}=0$ (no pressure mixing, empty squares), the predicted exponent is 
$1$. The numerical data agree well with the predictions close to the CP.
}
	\label{CriticalMidPointDiameter}
\end{figure}

Keeping our focus on the critical parent ($\nh=0$) we finally look at the midpoint diameter $\bar{n}$ vs.\ $\delta t$ (Fig.\ \ref{CriticalMidPointDiameter}), for which we have the theoretical prediction
Eqs.\ \eqref{diameter} and \eqref{diameterWPM} in the cases with and without pressure mixing. 
The leading order terms are seen clearly close to the CP. Notice in particular how without pressure mixing, the singular $\sim \delta t^{1-\alpha}$ dependence from the monodisperse case conspires with mixture effects to reproduce a rectilinear midpoint diameter, $|\nh|\sim \delta t$.

\section{Conclusions and Discussion}
\label{conc}

We have used complete scaling theory to relate standard 3D Ising criticality to polydisperse criticality, and thus to predict the scaling of a number of important properties of the phase diagram of polydisperse fluids. These predictions, which we summarise in Table \ref{TheTable},
were also confirmed in comparisons with numerical evaluations of the full theory.
We have emphasised the potential effects of pressure mixing in the scaling fields, and have highlighted a number of new observables that could be used to detect such effects.

\begin{table*}[ht]
\centering
\caption{Summary of analytical results (without prefactors). Abbreviations: `var.' = `variable', `diam.' = `diameter', `Frac.' = `Fractional', `vol.' = `volume', and `temp' = `temperature'. Remember that $\delta t_0\equiv t_{\rm cloud} - t_{\Delta=0}$.}
\begin{tabular}{|M{3.0cm}|M{1.7cm}|M{5.3cm}|M{5.3cm}|M{1.6cm}|l|}
	\hline
	Curve & Fixed var. & With pressure mixing & Without pressure mixing & \vspace{0.2cm} Illustration \\[5pt]
	\hline
	Cloud and others & $\Delta\neq0$ & 
	$t \sim \nh+
	\nh^2
	+\left|\nh\right|^{1/\bt}
	+\nh^3
	+\left|\nh\right|^{1/\beta}$ & $t \sim \nh+
	\nh^2
	+\left|\nh\right|^{1/\bt}
	+\nh^3
	+\left|\nh\right|^{1/\beta}$ & \vspace{0.2cm} Figs. \ref{CloudShadowPlot} \& \ref{Linear} \\[5pt]
	\hline
	Equal volume diam. & $\Delta=0$ & 
	$t \sim \nh
	+\nh^{1/(2\bt)}
	+\nh^{1/(2\beta)}$ & $t \sim \nh
	+\nh^{1/\left(1-\alpha\right)}$ & \vspace{0.2cm} Figs. \ref{Linear} \& \ref{ZeroDeltaFigure} \\[5pt]
	\hline
	$t_{\rm cloud} - t_{\Delta=0}$ vs.\ $\nh$ & $\Delta=\pm1$ \newline $\Delta=0$ & 
	$\delta t_0 \sim
	\nh^{1/(2\bt)}+
	\nh^{1/(2\beta)}$ 
	& $\delta t_0 \sim\nh~+~\nh^{1/(1-\alpha)}$
	& \vspace{0.4cm} Fig. \ref{CrossOverDifference} \\[10pt]
	\hline
	Shadow and others & $\Delta\neq0$ & 
	$t \sim \nh_{\pm}+
	\nh_{\pm}^2
	+\left|\nh_{\pm}\right|^{1/\bt}
	+\nh_{\pm}^3
	+\left|\nh_{\pm}\right|^{1/\beta}$ & $t \sim \nh_{\pm}+
	\nh_{\pm}^2
	+\left|\nh_{\pm}\right|^{1/\bt}
	+\nh_{\pm}^3
	+\left|\nh_{\pm}\right|^{1/\beta}$ & \vspace{0.2cm} Fig. \ref{CloudShadowPlot} \\[5pt]
	\hline
	$50$--$50$ coexistence & $\Delta=0$ & 
	$t \sim
	\nh_{\pm}^2
	+\left|\nh_{\pm}\right|^{1/\bt}
	+\nh_{\pm}^3
	+\left|\nh_{\pm}\right|^{1/\beta}$ & $t \sim
	\left|\nh_{\pm}\right|^{1/\bt}
	+\left|\nh_{\pm}\right|^{1/\beta}$ & \vspace{0.2cm} Fig. \ref{CoexCurveZeroDeltaAllfpm} \\[5pt]
	\hline
	Coexistence curve & $\nh=0$ & 
	$\delta t \sim
	\left|\nh_{\pm}\right|^{1/\bt}
	+\left|\nh_{\pm}\right|^{1/\beta}$ & $\delta t \sim
	\left|\nh_{\pm}\right|^{1/\bt}
	+\left|\nh_{\pm}\right|^{1/\beta}$ & \vspace{0.2cm} N/A \\[5pt]
	\hline
	Frac.\ vol.\ vs.\ temp.\ & $\nh=0$ & 
	$\Delta \sim
	\delta t^{\bt}$ & $\Delta \sim
	\delta t^{1-\bt}$ & \vspace{0.2cm} Fig. \ref{CriticalDeltaVSTemp} \\[5pt]
	\hline
	Midpoint diam. & $\nh=0$ & \parbox[t]{4cm}{$\delta t \sim \nbar^{1/(2\bt)}+\nbar^{1/(2\beta)}$ \\[0.4cm]
	$\bar n \sim \delta t^{2\bt}+\delta t^{2\bt+\alpha/\left(1-\alpha\right)}$} & \parbox[t]{4cm}{$\delta t \sim \nbar+\nbar^{1/(1-\alpha)}$  \\[0.4cm]
	 $\bar n \sim
	 \delta t+\delta t^{1/\left(1-\alpha\right)}$} & \vspace{0.6cm} Fig. \ref{CriticalMidPointDiameter} \\[15pt]
	\hline
\end{tabular}
		\label{TheTable}
	\end{table*}
A number of the potentially useful observables involve the equal volume diameter, which determines for any fixed overall (parent) density $\nh$ at what temperature a phase split with $\Delta=0$, i.e.\ with equal fractional volumes occupied by the coexisting phases, is produced. From Table \ref{TheTable} we see that this diameter itself has a leading linear variation independently of pressure mixing. The distance $\delta t_0$ to the cloud curve, which is the extra temperature decrease that is required to get from the onset of phase separation to a $50$--$50$ phase split, shows a clearer signature, with the leading density dependence being linear without pressure mixing but singular with exponent is $1/2\bt\simeq1.37$ when pressure mixing is present. The $50$--$50$ coexistence curve, which records the coexisting densities for parents on the equal volume diameter, is likewise a potentially useful probe: its leading quadratic term disappears without pressure mixing.

For measurements at fixed critical parent density, the difference $\Delta$ of the fractional phase volumes also shows clear signatures of pressure mixing, with a smaller exponent ($\bt$ rather than $1-\bt$) leading to significantly larger $\Delta$ near the critical point (CP) when pressure mixing is present. 
Finally, the midpoint diameter, defined as the average of the coexisting densities produced by a critical parent, is also sensitive to pressure mixing as found previously in the literature in analogous studies of monodisperse systems: the midpoint density $\nbar$ varying linearly with temperature without pressure mixing but singularly with exponent $2\bt$ when pressure mixing is present.

The above predictions are all amenable to experimental verification and as tools to detect pressure mixing effects. How easy this is will of course depend on the specific fluid system and in particular on the accuracy that can be achieved for measurements close to the CP. While we have found that some crossovers require quite a few orders of magnitude to see clearly, other properties like those connected to the equal fractional volume coexistence curve (Section \ref{fractional}) should be more easily accessible.
	
We note that in Ref.\ \citenum{ISI:000276971500024}, where a complete scaling theory for weakly compressible binary liquids was developed, the authors used literature data\cite{PhysRevA.28.1647} to show that the effective leading-order exponent of the midpoint diameter
is $0.75$, whereas the predicted leading and subleading exponents were respectively ${2\beta}\simeq0.65$ and ${1-\alpha}\simeq 0.89$. A possible explanation is that the data falls within a crossover range. Alternatively there could be impurities in the experimental set-up that
would act as a hidden field in the calculations, resulting in the Fisher-renormalisation\cite{PhysRev.176.257} of the leading exponent that we also find. This would change the exponent from $2\beta$ to $2\bt=2\beta/(1-\alpha)\simeq0.73$ (see Ref.\ \citenum{ISI:000254539700040}), close to the experimental result. Note that our results do not directly apply to the fixed pressure situation considered in Ref.~\citenum{ISI:000276971500024} since we have kept overall density fixed instead. (Fisher renormalisation of exponents at fixed composition and density was already found and discussed in Ref.\ \citenum{Anisimov1995} for the simpler revised scaling approach, in the context of binary mixtures.)

It is interesting to observe that if were dealing with mean-field criticality, where $\alpha=0$ and $\beta=\bt=1/2$, all terms that we have derived in the expansion of both the midpoint and equal volume diameters would simply degenerate to linear contributions, though -- as one might expect on general grounds due to fractionation -- the prefactors would differ between the two diameter definitions.

In future work we plan to look at additional properties of polydisperse colloidal fluids using the same setup presented here, i.e.\ a framework for polydisperse fluids with complete scaling, in the experimentally relevant case of controlled overall particle density.
Such properties could include Tolman's length,\cite{ISI:000243587200033} dielectric constant,\cite{ISI:000277265700029} refractive index,\cite{losada2012critical} thermal and transport properties,\cite{ISI:000276971500031,ISI:000337604100004,Anisimov1995} and more generally other properties that can be obtained from the dependence of the pressure on the other variables.\cite{behnejad2015isomorphic} It will be interesting to see what new physical insights can be gained from the inclusion of polydispersity and potential pressure mixing effects. We also plan to study more closely the crossover between monodisperse and polydisperse critical behaviour that one expects to see in weakly polydisperse systems, building on perturbative approaches to polydisperse critical behaviour.\cite{sollich2008weakly}

\begin{acknowledgments}
	PdC acknowledges financial support from CNPq, Conselho Nacional de Desenvolvimento Cient\'{i}fico e Tecnol\'{o}gico -- Brazil (GDE 202399/2014-1). PS acknowledges the stimulating research environment provided by the EPSRC Centre for Doctoral Training in Cross-Disciplinary Approaches to Non-Equilibrium Systems (CANES, EP/L015854/1). Both authors are grateful to Prof Nigel Wilding and to the other authors of Ref.\ \citenum{LJNigelPeter} for providing the simulation data (for a Lennard-Jones polydisperse fluid) which have been used in our numerical comparison.
\end{acknowledgments}

\appendix
\section{Expansion coefficients from mixing coefficients}
\label{CoeffsEqns}

As explained in the main text, by inserting Eqs.\ (\ref{nh_exp})--(\ref{muh_exp}) into Eqs.\ (\ref{cond0_mix})--(\ref{cond3_mix}), and comparing terms order by order,
we obtain sets of equations that involve no thermodynamic variables, i.e.\ they contain only coefficients. These can be solved to extract the expansion 
coefficients $\nu_i, \nu^\prime_i, \pi_i, \pi^\prime_i, \tau_i, \tau^\prime_i, \mv_i$ and $\mv^\prime_i$ in terms of the mixing coefficients. The conditions for the $\order(\Delta\rt)$ coefficients are
\bea
\label{firstorderconditions1}
0&=& \pi_1 - k_0 \tau_1 - \f\T\mv_1 \\
\label{firstorderconditions2}
0&=& -j_1\pi_1 + \tau_1 - \l_1\T\mv_1 \\
\label{firstorderconditions3}
0&=& -j_2\pi_1- k_2\tau_1 + \l_2\T\mv_1 \\
\label{firstorderconditions4}
-\lt_2 &=&  -\f\nu_1 + (2m_0\f+\n_3)\pi_1 + (n_0\f+\v_0)\tau_1
\nn
&&{} +
(\f\n_3\T+2\q_0)\mv_1
\eea
while for $\order(\rt^{2})$ 
one gets
\bea
\label{secondorderconditions1}
0&=& \pi_2 - k_0 \tau_2 - \f\T\mv_2 \\
\label{secondorderconditions2}
0&=& -j_1\pi_2 + \tau_2 - \l_1\T\mv_2 \\
\label{secondorderconditions3}
0&=& -j_2\pi_2- k_2\tau_2 + \l_2\T\mv_2 \\
\label{secondorderconditions4}
j_2\lt_2 &=&  -\f\nu_2 + (2m_0\f+\n_3)\pi_2 + (n_0\f+\v_0)\tau_2
\nn
&&{} +
(\f\n_3\T+2\q_0)\mv_2
\eea
The analogous relations for
the coefficients of the $\order(\rt^{1/\bt})$ terms read
\bea
0&=& \pi_3 - k_0 \tau_3 - \f\T\mv_3 \\
0&=& -j_1\pi_3 + \tau_3 - \l_1\T\mv_3 \\
0&=& -j_2\pi_3- k_2\tau_3 + \l_2\T\mv_3 \\
-b(j_1\f+\l_1)&=&  -\f\nu_3 + (2m_0\f+\n_3)\pi_3 + (n_0\f+\v_0)\tau_3 
\nn
&&{}+
(\f\n_3\T+2\q_0)\mv_3
\eea
and finally for $\order(\rt^{1/\beta})$:
\bea
0 &=& \pi_5 - k_0 \tau_5 - \f\T\mv_5 \\
-a &=& -j_1\pi_5 + \tau_5 - \l_1\T\mv_5 \\
0&=& -j_2\pi_5- k_2\tau_5 + \l_2\T\mv_5 \\
0 &=&\label{betapolydispersecondition} -\f\nu_5 + (2m_0\f+\n_3)\pi_5 + (n_0\f+\v_0)\tau_5
\nn
&&{}
+(\f\n_3\T+2\q_0)\mv_5.
\eea
These conditions all take the same form because they relate to terms that are {\em linear} in the `input variables' of the expansion described in the main text, i.e.\ $\Delta\rt$, $\rt^2$, $b\rt^{1/\bt}$ and $\rt^{1/\beta}$.
The $(\Delta\rt)^2$ and third order terms in the expansions (\ref{nh_exp})--(\ref{muh_exp}) are quadratic or third order in these input variables; the equations for their coefficients would therefore involve higher order mixing coefficients that we have not written down in Eqs.\ (\ref{pt_mix})--(\ref{ht_mix}), so we do not state them here.

The
conditions above are in principle
straightforward to solve, requiring only the inversion of the
$(M+3)\times(M+3)$ coefficient matrix on the right hand side. One expects that
all coefficients must generically be nonzero, although the
monodisperse case is exceptional (see Section \ref{mono}). Note that one can always proceed by first obtaining $\mv_i$ ($i=1,3,5$) from the last equation of each set to reduce the problem to three linear equations for the remaining expansion coefficients. This first step involves the inversion of the matrix $\f\n_3\T+2\q_0$. Except for $M=1$ this requires a nonzero (and in general full-rank, invertible) $\q_0$, supporting our conclusion in the main text that nonlinear mixing effects as specified by $\q_0$ need to be accounted for to obtain a meaningful description of critical mixture behaviour.

We note finally that the inhomogeneities on the left hand sides of (\ref{firstorderconditions1})---%
(\ref{firstorderconditions4}) and
(\ref{secondorderconditions1})---%
(\ref{secondorderconditions4}) are directly proportional to each other. Accordingly the expansion coefficients for $O(\Delta\rt)$ and $O(\rt^2)$ terms that these equations determine are likewise proportional:
\be
\nu_2=-j_2\nu_1, \mkern9mu\tau_2=-j_2\tau_1, \mkern9mu\pi_2=-j_2\pi_1,\mkern9mu \mv_2=-j_2\mv_1.
\label{j2relations}
\ee
From these results one deduces in particular that $\tilde\tau_2=0$ as explained in Section \ref{fractional} in the main text.

\section{Coexisting densities}
\label{MoreCoexDens}
In the following we present a number of additional analytical results (and intermediate steps) for the coexisting densities $\nh_\pm$ in both fixed-$\Delta$ and fixed-$\nh$ contexts. We start by writing down the result for $t$ vs.\ $\nh_\pm$ for fixed $\Delta$ (see Section \ref{fractional}):
\begin{widetext}
	\vspace{5cm}
	\bea
	t = &&{}\frac{\tau_1\Delta}{\pm\ltt+\Delta(\nu_1-\ltt)}\nh_\pm-
	\biggl[
	\frac{[\nu_2\pm g_2\Delta+(\nu_2'-g_2)\Delta^2]
		\tau_1\Delta}{\pm\ltt+
		\Delta(\nu_1-\ltt)}
	-\tau_2
	+\tau_2'\Delta^2\biggr]
	\left(\frac{\nh_\pm}{\pm\ltt+\Delta(\nu_1-\ltt)}\right)^2
	\nn
	& &{}
	-\Biggl(\frac{\nu_3\tau_1\Delta}{\pm\ltt+\Delta(\nu_1-\ltt)}-\tau_3\Biggr)\left(\frac{\nh_\pm}{\pm\ltt+\Delta(\nu_1-\ltt)}\right)^{1/\bt}
	- \Biggl\{
	\frac{[\pm g_4+(\nu_4-g_4)\Delta \pm g_4'\Delta^2+(\nu_4'-g_4')\Delta^3]
		\tau_1\Delta}{\pm\ltt+\Delta(\nu_1-\ltt)}-(\tau_4\Delta
	\nn
	&&{}
	+\tau_4'\Delta^3)- 2\left(\frac{\nu_2\pm g_2\Delta+(\nu_2'-g_2)\Delta^2}
	{\pm\ltt+\Delta(\nu_1-\ltt)}\right)
	\Biggl[
	\frac{[\nu_2\pm g_2\Delta+(\nu_2'-g_2)\Delta^2]\tau_1\Delta}
	{\pm\ltt+\Delta(\nu_1-\ltt)}
	- 
	(\tau_2+\tau_2'\Delta^2)\Biggr]\Biggr\}
	\left(\frac{\nh_\pm}{\pm\ltt+\Delta(\nu_1-\ltt)}\right)^3
	\nn
	& &{}-\left(\frac{\nu_5\tau_1\Delta}{\pm\ltt+\Delta(\nu_1-\ltt)}-\tau_5\right)
	\left(\frac{\nh_\pm}{\pm\ltt+\Delta(\nu_1-\ltt)}\right)^{1/\beta}.
	\eea
\end{widetext}
As pointed out in the main text, the shadow curve is obtained by taking $\nh_{\pm}$ for $\Delta=\mp1$, giving
\begin{widetext}
\bea
t &=& \frac{\tau_1}{\nu_1-2\ltt}\nh_\pm
-\Biggl[\frac{(\nu_2+\nu_2'-2g_2)\tau_1}{\nu_1-2\ltt}
-  (\tau_2+\tau_2')\Biggr]
\left(\frac{\nh_\pm}{\nu_1-2\ltt}\right)^2
-\left(\frac{\nu_3\tau_1}{\nu_1-2\ltt}-\tau_3\right)
\left|\frac{\nh_\pm}{\nu_1-2\ltt}\right|^{1/\bt}
\nn & &{}
- \Biggl\{
\frac{[\nu_4+\nu_4'-2(g_4+g_4')]\tau_1}{\nu_1-2\ltt}-(\tau_4+\tau_4')
- 2\left(\frac{\nu_2+\nu_2'-2g_2} {\nu_1-2\ltt}\right)
\Biggl[
\frac{(\nu_2+\nu_2' - 2g_2)\tau_1}{\nu_1-2\ltt}
- (\tau_2+\tau_2')\Biggr]\Biggr\}
\left(\frac{\nh_\pm}{\nu_1-2\ltt}\right)^3
\nn
& &{}-\left(\frac{\nu_5\tau_1}{\nu_1-2\ltt}-\tau_5\right)
\left|\frac{\nh_\pm}{\nu_1-2\ltt}\right|^{1/\beta}.
\label{shadow}
\eea
\end{widetext}

To obtain the coexisting densities at fixed $\nh$ (Section \ref{coexistence}) rather than fixed $\Delta$ as above, one starts from the general expansion~\eqref{nh_pm2} and then uses Eq.~\eqref{Delta_rt_soln} to eliminate $\Delta\rt$, to obtain an expansion in terms of $\rt$ only:
\begin{widetext}
\bea
\nh_\pm &=& \nh \pm \ltt\rt - \ltt\Delta\rt
\pm g_2\rt(\Delta\rt) - g_2(\Delta\rt^2)
\pm g_4\rt^3 -g_4\rt^2(\Delta\rt)
\pm g_4'\rt(\Delta\rt)^2 -g_4'(\Delta\rt)^3
\\
&=&{} \left(1-\frac{\ltt}{\nu_1}\right)\nh 
\pm \ltt\rt 
+ \frac{\nu_2\ltt}{\nu_1}\rt^2
\pm g_2\rt\frac{\nh}{\nu_1}
+ \left(\frac{\nu_2'\ltt}{\nu_1}-g_2\right) \left(\frac{\nh}{\nu_1}\right)^2+\frac{\nu_3\ltt}{\nu_1}\rt^{1/\bt}
\pm\left(g_4-\frac{\nu_2g_2}{\nu_1}\right)\rt^3
\nn
&&{}
+\Biggl[-g_4 + \frac{\nu_4\ltt}{\nu_1}+2\frac{\nu_2}{\nu_1}\Biggl(g_2
-\frac{\nu_2'\ltt}{\nu_1}\Biggr)
\Biggr]
\rt^2\frac{\nh}{\nu_1}
\pm\left(g_4'-\frac{\nu_2'g_2}{\nu_1}\right)\rt
\left(\frac{\nh}{\nu_1}\right)^2
\nn
& &{}
+\left[-g_4'+\frac{\nu_4'\ltt}{\nu_1}+2\frac{\nu_2'}{\nu_1}
\left(g_2-\frac{\nu_2'\ltt}{\nu_1}\right)\right]
\left(\frac{\nh}{\nu_1}\right)^3
+\frac{\nu_5\ltt}{\nu_1}\rt^{1/\beta}.
\label{nhpm_coex}
\eea
\end{widetext}
As explained in the main text, one can then determine $\rt$ from this expression to eliminate it from a similar expansion for the temperature, Eq.~\eqref{deltat}, in order to determine the temperature dependence of the coexisting densities $\nh_\pm$ at fixed $\nh$ that is stated in Eq.~\eqref{CoexCurveFixedDens}.
For the diameter, defined as $\nbar\equiv\half(\nh\one+\nh\two)$, one can proceed similarly. Starting again from Eq.~\eqref{nh_pm2} and using \eqref{parentvsrt}
\bea
\nbar &=& \nh -\Delta[\ltt\rt + g_2\Delta\rt^2+g_4\rt^3+g_4'\Delta^2\rt^3]
\label{nbar_general}
\\
&=& (\nu_1-\ltt)\Delta\rt +(\nu_2-g_2\Delta+\nu_2'\Delta^2)\rt^2+
\nu_3\rt^{1/\bt}
\nn
& &{}+ (g_4+\nu_4\Delta +
g_4'\Delta^2+\nu_4'\Delta^3)\rt^3 + \nu_5\rt^{1/\beta}.
\eea
Once $\Delta$ is eliminated to fix $\nh$, again with \eqref{Delta_rt_soln}, we have
\bea
\nbar &=& 
\left(1-\frac{\ltt}{\nu_1}\right) \nh
+ \left(\frac{\nu_2'\ltt}{\nu_1}-g_2\right) \left(\frac{\nh}{\nu_1}\right)^2
+ \frac{\nu_2\ltt}{\nu_1}\rt^2
\nn
& &{}+\frac{\nu_3\ltt}{\nu_1}\rt^{1/\bt}+\frac{\nh}{\nu_1}\rt^2\left[\frac{2j_2\ltt\nu_2'}{\nu_1}+\frac{\ltt{\nu_4}}{\nu_1}-2j_2g_2-g_4\right]
\nn
& &{}+\left(\frac{\nh}{\nu_1}\right)^3\left[\frac{2g_2\nu_2'}{\nu_1}-\frac{2\ltt{\nu_2'}^{2}}{\nu_1^2}+\frac{\ltt\nu_4'}{\nu_1}-g_4'\right]
\nn
& &{}+\frac{\nu_5\ltt}{\nu_1}\rt^{1/\beta}
\label{nbar_coex}
\eea
Again one has to eliminate $\rt$. For the diameter it is convenient to have an expression for $\nbar$ in terms of $\delta t$ rather than vice versa, so here one solves Eq.~\eqref{deltat} for $\rt$. Focussing on the simplest case $\nh=0$, this yields
\bea
\rt&=&
\left(\frac{\delta t}{\tilde\tau_3}\right)^{\bt} -
\bt\frac{\tilde\tau_5}{\tilde\tau_3}\left(\frac{\delta t}{\tilde\tau_3}\right)^{\bt+\alpha/\left(1-\alpha\right)}.
\label{rt_elimination}
\eea
and after insertion into Eq.~(\ref{nbar_coex}) one finds \eqref{diameter}.
When pressure mixing is absent, one can check by going back to Eq.~\eqref{Delta_rt_soln} that, again for $\nh=0$,  all terms with 
integer powers vanish in Eq.~(\ref{nbar_coex}):
\be
\nbar = \frac{\ltt}{\nu_1}\left(\nu_3\rt^{1/\bt}+\nu_5\rt^{1/\beta}\right).
\ee
Eliminating $\rt$ again using Eq.~(\ref{rt_elimination}) then leads to
Eq.~\eqref{diameterWPM} in the main text.

\section{Equivalence between mixtures with different numbers of species}
\label{equivalence}
	
In this appendix we explain how the critical phase behaviour of a fluid mixture with a certain number of species can be described equivalently in terms of a mixture with an `inflated' number of species. As explained in the main text, this is useful in order to understand how the monodisperse limit can be approached starting from a polydisperse system.

Consider a mixture with $M^\prime$ species and mixing coefficients $j_1^\prime,j_2^\prime,k_0^\prime,k_2^\prime,\l_0^\prime, \l_1^\prime, \l_2^{\prime},\bm{q}_0^{\prime}$, etc. We want to map it onto a system with $M> M^\prime$ and mixing coefficients $j_1,j_2,k_0,k_2,\l_0, \l_1, \l_2,\bm{q}_0$, etc.\ such that both systems are physically equivalent with respect to their phase behaviour near the critical point. We nominally create new species by `painting' particles, allowing them to be distinguished by an additional colour label without changing any of their physical properties. Each original species is thus divided into one or more coloured subspecies, in such a way that
\be
\rho_i^\prime=\sum\limits_{j\in i}\rho_j
\label{cons_colouring}
\ee
where $\rho_i^\prime$ is the density of a particle species $i$ in the original labelling and the $\rho_j$ are similarly the densities of the subspecies arising from $i$ by colouring, as indicated symbolically in the sum by the notation $j\in i$.
In terms of the new, $M$-species description, the free energy density of the system is (setting $k_B=1$ here)
\be
f=T\sum\limits_{j=1}^{M}\rho_j(\ln \rho_j - 1)+f_{\text ex}(\{\rho_i^\prime\})
\ee
where $f_{\text ex}$ is the excess free energy density and can be expressed as a function of the original composition $\{\rho_i^\prime\}$ since colouring the particles by definition does not affect their physical interactions.
	
The chemical potentials of the coloured particles are obtained from $f$ by differentiation as
\be
\mu_j = \frac{\partial f}{\partial \rho_j}=
T\ln\rho_j+\frac{\partial f_{\text ex}(\{\rho_i^\prime\})}{\partial \rho_j}
\ee
Now because of Eq.~(\ref{cons_colouring}), changing $\rho_j$ changes only the density $\rho_i'$ of the original species that subspecies $j$ belongs to, by the same amount. 
It follows that
\be
\mu_j - T\ln\rho_j = \frac{\partial f_{\text ex}}{\partial \rho_i'} \qquad \mbox{for\ all\ }j\in i
\label{excess_chem_pot_j}
\ee
The intuitive content of this relation is that the excess chemical potential, which encodes the physical properties of each particle type, is the same for subspecies $j$ as for the original species $i$ that was coloured to obtain $j$.

Now consider a small change $d\mu_j$ in all subspecies chemical potentials. Eq.~(\ref{excess_chem_pot_j}) then implies
\be
d\mu_j = T\frac{d\rho_j}{\rho_j} + 
\sum_k \frac{\partial^2 f_{\text ex}}{\partial \rho_i'\partial \rho_k'}\, d\rho_k'
\ee
Multiplying by $\rho_j$ and summing over $j\in i$ gives, using on the r.h.s.\ again~(\ref{cons_colouring}),
\be
\sum_{j\in i} \rho_j\,d\mu_j = T d\rho_i' + 
\sum_k \rho_i'\,\frac{\partial^2 f_{\text ex}}{\partial \rho_i'\partial \rho_k'}\, d\rho_k'
\ee
This can be read as a vector equation for the density changes $d\rho_i'$ of the original species as a function of the weighted chemical potential changes associated with each original species, $\sum_{j\in i} \rho_j\,d\mu_j$, on the l.h.s. In particular, if these weighted changes are all zero, then also the original species densities remain unchanged, $d\rho_i' \equiv 0$. As we are keeping temperature $T$ fixed, also pressure must then remain unchanged. In summary, any chemical potential change that obeys
\be
\sum\limits_{j\in i}\rho_j\,d\mu_j=0
\label{SpeciesGroupSum}
\ee 
for all $i$ will not change the physical state of the system in any way. (What it will modify is the physically irrelevant distribution of coloured subspecies within each physical particle species.)
In the mapping equations (\ref{pt_mix})--(\ref{ht_mix}), such a chemical potential change must then leave the Ising scaling variables on the l.h.s.\ unchanged, as well as the temperature $t$ and pressure $\ph$ on the r.h.s.

The above invariance with respect to most changes of the chemical potentials for the labelled particle species puts significant constraints on the mixing coefficients for the $M$-species system in Eqs.~(\ref{pt_mix})--(\ref{ht_mix}) as we now show. Consider first the simplest case $M'=1$. In vector form, the chemical potential changes we are considering obey $\rh \T d\muh=0$; we can use the shifted and scaled chemical potentials $\muh$ here because their changes are proportional to those of the conventional chemical potentials. For any such $d\muh$ the Ising scaling pressure must not change, hence $d\pt=(\partial \pt/\partial \muh)\T d\muh=0$. This implies that the vectors $\rh$ and $\partial \pt/\partial \muh$ are proportional as otherwise one could find a $d\muh$ that is orthogonal to $\rh$ but not to $\partial \pt/\partial \muh$. Applying the same logic to the other two scaling variables shows that
\bea 
\rh \propto \frac{\partial \pt}{\partial \muh} \propto  \frac{\partial \tt}{\partial \muh}
\propto  \frac{\partial \htt}{\partial \muh}
\label{grad_prop}
\eea
In other words, at any state point the gradients of the three mapping equations w.r.t.\ the chemical potentials must be proportional to each other. Writing out the gradient for e.g.\ $\pt$ from Eq.~(\ref{pt_mix}) one has
\bea
\frac{\partial\pt}{\partial \muh} =
- \l_0 - 2\q_0\muh - t \v_0 - \ph \n_3
\label{pt_grad}
\eea
At the critical point this reduces to $-\l_0$, and comparing with the analogous gradients for $\tt$ and $\htt$ shows that
\bea 
\l_0 \propto \l_1 \propto \l_2
\eea
As $\l_0=\f$, also $\l_1$ and $\l_2$ must then be proportional to $\f$, and it remains to fix their normalisation. This can be done by comparing Eq.~(\ref{rh_gen}) for the densities $\rh$ in the coloured ($M$-species) system with the analogous expression for the density in the original $M'=1$ system. Using the constraint that the labelled densities must add up to the original density, Eq.~(\ref{cons_colouring}), then shows that we must have $\ev\T \l_1 = l'_1$ and $\ev\T\l_2=l'_2$. Since $\ev\T\f=1$ by construction, the explicit mapping from $M'=1$ to an equivalent $M$-species system for these mixing coefficients is
\bea 
\l_1 = \f l_1', \qquad \l_2 = \f l_2'
\eea
By comparing gradients like (\ref{pt_grad}) at nonzero $t$ or $\ph$ one shows easily that also $\v_0=\f v_0'$, $\n_3=\f n_3'$ with analogous results for all other vector mixing coefficients. For the matrix $\q_0$ one notes that $\q_0\muh$ must also be proportional to $\f$, otherwise one could change the direction of the gradient (\ref{pt_grad}) by moving around state space and do so in a different way for the different scaling variables, thus destroying the proportionality~(\ref{grad_prop}). By symmetry of $\q_0$ and the same normalisation argument as above one then finds
\bea 
\q_0 = \f q_0' \f\T
\eea
in terms of the mixing coefficient $q_0'$ of the original $M^{\prime}=1$ system. Analogous results apply for the other matrix mixing coefficients.

The generalisation of the above reasoning to $M^{\prime}>1$ is not difficult. If we define for each $i$ the vector $\rh_i$ as the one collecting the densities of the corresponding labelled subspecies $j\in i$, with all other entries set to zero, then the chemical potential changes that leave the physical state of the system invariant obey $\rh_i \T d\muh=0$ for all $i$. The three scaling variable gradients [$(\partial \pt/\partial \muh)$, etc.] must then be linear combinations of the vectors $\rh_i$ at each state point. At the critical point these vectors are proportional to the vectors $\f_i$, defined such that $\f_i$ collects the nonzero entries of $\f$ that correspond to original species $i$. One thus has, taking $\l_1$ as an example
\bea 
\l_1 = \sum_i L_i \f_i
\label{l1_transform}
\eea
The coefficients $L_i$ can be worked out from the density sum constraints (\ref{cons_colouring}) again, giving
\bea 
\l_1 = \sum_i \f_i \,(\l'_1)_i / f'_i
\eea
where the parent distribution in the original system, $f'_i = \ev\T \f_i=\sum_{j\in i} f_j$, naturally emerges as normaliser. If we define a rectangular $M\times M'$ `inflation' matrix by
\bea 
S_{ji} = f_j / f'_i \qquad \mbox{for\ }j\in i
\eea
and $S_{ji}=0$ otherwise, then the relation~(\ref{l1_transform}) can be written in the simple matrix form
\bea
\l_1 = \bm{S} \, \l_1'
\eea
This then applies to all vector mixing coefficients. Matrix mixing coefficients are inflated from $M'$ to $M$ similarly by
\bea 
\q_0 = \bm{S} \,\q_0'\, \bm{S}\T
\eea
etc. This transformation would then generalise to tensorial mixing coefficients in the obvious way, which would arise if one chose to include third or higher order terms in the mapping equations (\ref{pt_mix})--(\ref{ht_mix}). In essence the transformation ensures that the $M$ chemical potentials $\muh$ in the larger system enter all physical properties only through the $M'$ weighted combinations $\bm{S}\T\muh$, consistent with the invariance condition (\ref{SpeciesGroupSum}). 

Beyond conceptual use in understanding the monodisperse limit, the above method could also be deployed in fitting mixing coefficients to numerical data for the critical behaviour of polydisperse systems. In principle a description with $M$ as large as possible is preferred as it would capture the most detail; on the other hand,
the computational costs will increase with the number of fitting parameters and hence with $M$ (see next Section). One could therefore envisage initially fitting parameters for some small $M'$ and then using the method above to map these to an equivalent representation with larger $M$. This should then constitute a suitable initial point for further parameter optimisation in the larger description. The process could be repeated in an iterative manner, building up increasingly refined data-driven descriptions of the critical behaviour of a polydisperse system.

\section{Fitted mixing coefficients}
\label{AppNum}
In order to find fitted mixing coefficients, we first created a forward routine that, for any given set of mixing coefficients, solves the full system of equations (mapping equations and dilution line constraint) numerically to produce predictions for the cloud and shadow curves. The root mean squared error between these predictions and those Lennard-Jones data points from Ref.~\citenum{LJNigelPeter} that were not too far from the CP was 
then used as the objective function, to be minimised across all possible assignments of mixing coefficients. Only $\l_0$ is taken as fixed because as shown in the main text it has to equal the normalised parent distribution $\f$.

Even for the smallest mixture description, $M=2$, this still leaves 38 linear and quadratic mixing coefficients to be found; for general $M$ this number is $\left(26 + 19 M + 3 M^2\right)/2$, bearing in mind that each mixing vector contributes $M$ parameters and each (symmetric) mixing matrix $M(M+1)/2$.
Minimisation of the root mean square objective over such a large parameter space proved difficult, in particular due to the presence of a large number of local minima.
We therefore chose to simplify the set of mixing coefficients by keeping all first order coefficients plus $\q_0$, and setting the remaining coefficients of quadratic terms to zero. The choice of $\q_0$ was driven by the fact that,
as explained in Appendix \ref{CoeffsEqns}, the presence of this term is essential in order to have a consistent theory that can properly describe fractionation.
Of course we expect that in reality the other quadratic mixing coefficients will not be exactly zero but, as shown in Fig.~\ref{CloudShadowPlot}, setting them to zero as we did is reasonable given that it still allows us to fit the simulation data well. For the same reason we fixed the amplitude parameter in the scaling relation \eqref{pt_scaling_form} to $Q=1$.

Given the considerations above we limited ourselves to fitting mixing coefficients for a mixture description with the smallest non-trivial value, $M=2$. One then has to make an appropriate assignment of the effective parent distribution $\f$ for this chosen $M$. The actual distribution of the polydisperse attributed used in Ref.\ \citenum{LJNigelPeter} was roughly bell-shaped, more specifically a Schulz distribution with a standard deviation of $14$\% of the mean, limited
to the range $0.5<\sigma<1.4$ and then renormalised appropriately.
To find $\f$ we formed $M$ bins evenly spaced across the $\sigma$-range and then integrated the probability within each bin.
For $M=2$ this led to 
\bea
\f\T=\begin{bmatrix}
0.378 & 0.622\end{bmatrix}
\eea
The best mixing coefficients corresponding to those choice of $\f$ were found using a combination of global optimisation using simulated annealing and local optimisation near candidate local minima. This produced the following nonzero coefficients:
\bea
\bm{l}_1\T&=&\begin{bmatrix}
	-0.836 & -0.987
\end{bmatrix}
\eea
\bea
\bm{l}_2\T&=&\begin{bmatrix}
	1.10 & 1.05
\end{bmatrix}
\eea
\bea
\bm{q}_0&=
\begin{bmatrix} 
	0.980 & 0.147\\
	0.147 & 0.600
\end{bmatrix}
\label{q0}
\eea
\bea
j_1=0.231,j_2=0.217
\eea
\bea
k_0= -0.599,k_2=-0.996
\eea

We also considered an alternative route towards fitting mixing coefficients, where cloud and shadow curve data are first fitted to the form we find in our analytical expansions [Eqs.~\eqref{cloud} and \eqref{shadow}], using arbitrary prefactors. Given our theoretical predictions that express these prefactors in terms of the mixing coefficients one could then, in a second step,
find a set of coefficients that reproduces the fitted prefactors. While this approach is a priori attractive, it proved to be computationally no simpler and also has two conceptual drawbacks. Firstly, as our analytical expansions are truncated beyond a certain order, there is a possibility that they would be accidentally used for fitting in a region where the discarded terms would be significant. 
Secondly, using the theoretically predicted expansions for cloud and shadow as part of the fitting procedure would have removed the possibility of testing these predictions in an unbiased manner.

\bibliography{PolyCriticalBib}

\begin{thebibliography}{52}%
\makeatletter
\providecommand \@ifxundefined [1]{%
 \@ifx{#1\undefined}
}%
\providecommand \@ifnum [1]{%
 \ifnum #1\expandafter \@firstoftwo
 \else \expandafter \@secondoftwo
 \fi
}%
\providecommand \@ifx [1]{%
 \ifx #1\expandafter \@firstoftwo
 \else \expandafter \@secondoftwo
 \fi
}%
\providecommand \natexlab [1]{#1}%
\providecommand \enquote  [1]{``#1''}%
\providecommand \bibnamefont  [1]{#1}%
\providecommand \bibfnamefont [1]{#1}%
\providecommand \citenamefont [1]{#1}%
\providecommand \href@noop [0]{\@secondoftwo}%
\providecommand \href [0]{\begingroup \@sanitize@url \@href}%
\providecommand \@href[1]{\@@startlink{#1}\@@href}%
\providecommand \@@href[1]{\endgroup#1\@@endlink}%
\providecommand \@sanitize@url [0]{\catcode `\\12\catcode `\$12\catcode
  `\&12\catcode `\#12\catcode `\^12\catcode `\_12\catcode `\%12\relax}%
\providecommand \@@startlink[1]{}%
\providecommand \@@endlink[0]{}%
\providecommand \url  [0]{\begingroup\@sanitize@url \@url }%
\providecommand \@url [1]{\endgroup\@href {#1}{\urlprefix }}%
\providecommand \urlprefix  [0]{URL }%
\providecommand \Eprint [0]{\href }%
\providecommand \doibase [0]{http://dx.doi.org/}%
\providecommand \selectlanguage [0]{\@gobble}%
\providecommand \bibinfo  [0]{\@secondoftwo}%
\providecommand \bibfield  [0]{\@secondoftwo}%
\providecommand \translation [1]{[#1]}%
\providecommand \BibitemOpen [0]{}%
\providecommand \bibitemStop [0]{}%
\providecommand \bibitemNoStop [0]{.\EOS\space}%
\providecommand \EOS [0]{\spacefactor3000\relax}%
\providecommand \BibitemShut  [1]{\csname bibitem#1\endcsname}%
\let\auto@bib@innerbib\@empty
\bibitem [{\citenamefont {Vi{\'e}ville}, \citenamefont {Tanty},\ and\
  \citenamefont {Delsuc}(2011)}]{vieville2011polydispersity}%
  \BibitemOpen
  \bibfield  {author} {\bibinfo {author} {\bibfnamefont {J.}~\bibnamefont
  {Vi{\'e}ville}}, \bibinfo {author} {\bibfnamefont {M.}~\bibnamefont {Tanty}},
  \ and\ \bibinfo {author} {\bibfnamefont {M.-A.}\ \bibnamefont {Delsuc}},\
  }\bibfield  {title} {\enquote {\bibinfo {title} {Polydispersity index of
  polymers revealed by dosy nmr},}\ }\href@noop {} {\bibfield  {journal}
  {\bibinfo  {journal} {Journal of Magnetic Resonance}\ }\textbf {\bibinfo
  {volume} {212}},\ \bibinfo {pages} {169--173} (\bibinfo {year}
  {2011})}\BibitemShut {NoStop}%
\bibitem [{\citenamefont {Auer}\ and\ \citenamefont
  {Frenkel}(2001)}]{auer2001suppression}%
  \BibitemOpen
  \bibfield  {author} {\bibinfo {author} {\bibfnamefont {S.}~\bibnamefont
  {Auer}}\ and\ \bibinfo {author} {\bibfnamefont {D.}~\bibnamefont {Frenkel}},\
  }\bibfield  {title} {\enquote {\bibinfo {title} {Suppression of crystal
  nucleation in polydisperse colloids due to increase of the surface free
  energy},}\ }\href@noop {} {\bibfield  {journal} {\bibinfo  {journal}
  {Nature}\ }\textbf {\bibinfo {volume} {413}},\ \bibinfo {pages} {711--713}
  (\bibinfo {year} {2001})}\BibitemShut {NoStop}%
\bibitem [{\citenamefont {Belli}\ \emph {et~al.}(2011)\citenamefont {Belli},
  \citenamefont {Patti}, \citenamefont {Dijkstra},\ and\ \citenamefont
  {Van~Roij}}]{belli2011polydispersity}%
  \BibitemOpen
  \bibfield  {author} {\bibinfo {author} {\bibfnamefont {S.}~\bibnamefont
  {Belli}}, \bibinfo {author} {\bibfnamefont {A.}~\bibnamefont {Patti}},
  \bibinfo {author} {\bibfnamefont {M.}~\bibnamefont {Dijkstra}}, \ and\
  \bibinfo {author} {\bibfnamefont {R.}~\bibnamefont {Van~Roij}},\ }\bibfield
  {title} {\enquote {\bibinfo {title} {Polydispersity stabilizes biaxial
  nematic liquid crystals},}\ }\href@noop {} {\bibfield  {journal} {\bibinfo
  {journal} {Physical Review Letters}\ }\textbf {\bibinfo {volume} {107}},\
  \bibinfo {pages} {148303} (\bibinfo {year} {2011})}\BibitemShut {NoStop}%
\bibitem [{\citenamefont {Sollich}(2005)}]{sollich2005nematic}%
  \BibitemOpen
  \bibfield  {author} {\bibinfo {author} {\bibfnamefont {P.}~\bibnamefont
  {Sollich}},\ }\bibfield  {title} {\enquote {\bibinfo {title} {Nematic-nematic
  demixing in polydisperse thermotropic liquid crystals},}\ }\href@noop {}
  {\bibfield  {journal} {\bibinfo  {journal} {The Journal of Chemical Physics}\
  }\textbf {\bibinfo {volume} {122}},\ \bibinfo {pages} {214911} (\bibinfo
  {year} {2005})}\BibitemShut {NoStop}%
\bibitem [{\citenamefont {Rogosi\'c}, \citenamefont {Mencer},\ and\
  \citenamefont {Gomzi}(1996)}]{ROGOSIC19961337}%
  \BibitemOpen
  \bibfield  {author} {\bibinfo {author} {\bibfnamefont {M.}~\bibnamefont
  {Rogosi\'c}}, \bibinfo {author} {\bibfnamefont {H.}~\bibnamefont {Mencer}}, \
  and\ \bibinfo {author} {\bibfnamefont {Z.}~\bibnamefont {Gomzi}},\ }\bibfield
   {title} {\enquote {\bibinfo {title} {Polydispersity index and molecular
  weight distributions of polymers},}\ }\href {\doibase
  http://dx.doi.org/10.1016/S0014-3057(96)00091-2} {\bibfield  {journal}
  {\bibinfo  {journal} {European Polymer Journal}\ }\textbf {\bibinfo {volume}
  {32}},\ \bibinfo {pages} {1337 -- 1344} (\bibinfo {year} {1996})}\BibitemShut
  {NoStop}%
\bibitem [{\citenamefont {Poon}(2002)}]{poon2002physics}%
  \BibitemOpen
  \bibfield  {author} {\bibinfo {author} {\bibfnamefont {W.}~\bibnamefont
  {Poon}},\ }\bibfield  {title} {\enquote {\bibinfo {title} {The physics of a
  model colloid--polymer mixture},}\ }\href@noop {} {\bibfield  {journal}
  {\bibinfo  {journal} {Journal of Physics: Condensed Matter}\ }\textbf
  {\bibinfo {volume} {14}},\ \bibinfo {pages} {R859} (\bibinfo {year}
  {2002})}\BibitemShut {NoStop}%
\bibitem [{\citenamefont {Van~den Pol}\ \emph {et~al.}(2009)\citenamefont
  {Van~den Pol}, \citenamefont {Petukhov}, \citenamefont {Thies-Weesie},
  \citenamefont {Byelov},\ and\ \citenamefont {Vroege}}]{van2009experimental}%
  \BibitemOpen
  \bibfield  {author} {\bibinfo {author} {\bibfnamefont {E.}~\bibnamefont
  {Van~den Pol}}, \bibinfo {author} {\bibfnamefont {A.}~\bibnamefont
  {Petukhov}}, \bibinfo {author} {\bibfnamefont {D.}~\bibnamefont
  {Thies-Weesie}}, \bibinfo {author} {\bibfnamefont {D.}~\bibnamefont
  {Byelov}}, \ and\ \bibinfo {author} {\bibfnamefont {G.}~\bibnamefont
  {Vroege}},\ }\bibfield  {title} {\enquote {\bibinfo {title} {Experimental
  realization of biaxial liquid crystal phases in colloidal dispersions of
  boardlike particles},}\ }\href@noop {} {\bibfield  {journal} {\bibinfo
  {journal} {Physical Review Letters}\ }\textbf {\bibinfo {volume} {103}},\
  \bibinfo {pages} {258301} (\bibinfo {year} {2009})}\BibitemShut {NoStop}%
\bibitem [{\citenamefont {Stuart}, \citenamefont {Scheutjens},\ and\
  \citenamefont {Fleer}(1980)}]{stuart1980polydispersity}%
  \BibitemOpen
  \bibfield  {author} {\bibinfo {author} {\bibfnamefont {M.}~\bibnamefont
  {Stuart}}, \bibinfo {author} {\bibfnamefont {J.~M.~H.}\ \bibnamefont
  {Scheutjens}}, \ and\ \bibinfo {author} {\bibfnamefont {G.}~\bibnamefont
  {Fleer}},\ }\bibfield  {title} {\enquote {\bibinfo {title} {Polydispersity
  effects and the interpretation of polymer adsorption isotherms},}\
  }\href@noop {} {\bibfield  {journal} {\bibinfo  {journal} {Journal of Polymer
  Science: Polymer Physics Edition}\ }\textbf {\bibinfo {volume} {18}},\
  \bibinfo {pages} {559--573} (\bibinfo {year} {1980})}\BibitemShut {NoStop}%
\bibitem [{\citenamefont {Evans}, \citenamefont {Fairhurst},\ and\
  \citenamefont {Poon}(1998)}]{evans1998universal}%
  \BibitemOpen
  \bibfield  {author} {\bibinfo {author} {\bibfnamefont {R.}~\bibnamefont
  {Evans}}, \bibinfo {author} {\bibfnamefont {D.}~\bibnamefont {Fairhurst}}, \
  and\ \bibinfo {author} {\bibfnamefont {W.}~\bibnamefont {Poon}},\ }\bibfield
  {title} {\enquote {\bibinfo {title} {Universal law of fractionation for
  slightly polydisperse systems},}\ }\href@noop {} {\bibfield  {journal}
  {\bibinfo  {journal} {Physical Review Letters}\ }\textbf {\bibinfo {volume}
  {81}},\ \bibinfo {pages} {1326} (\bibinfo {year} {1998})}\BibitemShut
  {NoStop}%
\bibitem [{\citenamefont {Wilding}, \citenamefont {Sollich},\ and\
  \citenamefont {Buzzacchi}(2008)}]{PhysRevE.77.011501}%
  \BibitemOpen
  \bibfield  {author} {\bibinfo {author} {\bibfnamefont {N.~B.}\ \bibnamefont
  {Wilding}}, \bibinfo {author} {\bibfnamefont {P.}~\bibnamefont {Sollich}}, \
  and\ \bibinfo {author} {\bibfnamefont {M.}~\bibnamefont {Buzzacchi}},\
  }\bibfield  {title} {\enquote {\bibinfo {title} {Polydisperse lattice-gas
  model},}\ }\href {\doibase 10.1103/PhysRevE.77.011501} {\bibfield  {journal}
  {\bibinfo  {journal} {Physical Review E}\ }\textbf {\bibinfo {volume} {77}},\
  \bibinfo {pages} {011501} (\bibinfo {year} {2008})}\BibitemShut {NoStop}%
\bibitem [{\citenamefont {de~Castro}\ and\ \citenamefont
  {Sollich}(2017)}]{PabloPeter1}%
  \BibitemOpen
  \bibfield  {author} {\bibinfo {author} {\bibfnamefont {P.}~\bibnamefont
  {de~Castro}}\ and\ \bibinfo {author} {\bibfnamefont {P.}~\bibnamefont
  {Sollich}},\ }\bibfield  {title} {\enquote {\bibinfo {title} {Phase
  separation dynamics of polydisperse colloids: a mean-field lattice-gas
  theory},}\ }\href {\doibase 10.1039/C7CP04062H} {\bibfield  {journal}
  {\bibinfo  {journal} {Physical Chemistry Chemical Physics}\ }\textbf
  {\bibinfo {volume} {19}},\ \bibinfo {pages} {22509--22527} (\bibinfo {year}
  {2017})}\BibitemShut {NoStop}%
\bibitem [{\citenamefont {Berger}, \citenamefont {Lucas},\ and\ \citenamefont
  {Twersky}(1991)}]{berger1991polydisperse}%
  \BibitemOpen
  \bibfield  {author} {\bibinfo {author} {\bibfnamefont {N.}~\bibnamefont
  {Berger}}, \bibinfo {author} {\bibfnamefont {R.}~\bibnamefont {Lucas}}, \
  and\ \bibinfo {author} {\bibfnamefont {V.}~\bibnamefont {Twersky}},\
  }\bibfield  {title} {\enquote {\bibinfo {title} {Polydisperse scattering
  theory and comparisons with data for red blood cells},}\ }\href@noop {}
  {\bibfield  {journal} {\bibinfo  {journal} {The Journal of the Acoustical
  Society of America}\ }\textbf {\bibinfo {volume} {89}},\ \bibinfo {pages}
  {1394--1401} (\bibinfo {year} {1991})}\BibitemShut {NoStop}%
\bibitem [{\citenamefont {Jos{\'e}-Yacam{\'a}n}\ \emph
  {et~al.}(1996)\citenamefont {Jos{\'e}-Yacam{\'a}n}, \citenamefont
  {Rend{\'o}n}, \citenamefont {Arenas},\ and\ \citenamefont
  {Puche}}]{jose1996maya}%
  \BibitemOpen
  \bibfield  {author} {\bibinfo {author} {\bibfnamefont {M.}~\bibnamefont
  {Jos{\'e}-Yacam{\'a}n}}, \bibinfo {author} {\bibfnamefont {L.}~\bibnamefont
  {Rend{\'o}n}}, \bibinfo {author} {\bibfnamefont {J.}~\bibnamefont {Arenas}},
  \ and\ \bibinfo {author} {\bibfnamefont {M.~C.~S.}\ \bibnamefont {Puche}},\
  }\bibfield  {title} {\enquote {\bibinfo {title} {Maya blue paint: an ancient
  nanostructured material},}\ }\href@noop {} {\bibfield  {journal} {\bibinfo
  {journal} {Science}\ }\textbf {\bibinfo {volume} {273}},\ \bibinfo {pages}
  {223} (\bibinfo {year} {1996})}\BibitemShut {NoStop}%
\bibitem [{\citenamefont {Anema}(2008)}]{anema2008effect}%
  \BibitemOpen
  \bibfield  {author} {\bibinfo {author} {\bibfnamefont {S.~G.}\ \bibnamefont
  {Anema}},\ }\bibfield  {title} {\enquote {\bibinfo {title} {Effect of milk
  solids concentration on whey protein denaturation, particle size changes and
  solubilization of casein in high-pressure-treated skim milk},}\ }\href@noop
  {} {\bibfield  {journal} {\bibinfo  {journal} {International Dairy Journal}\
  }\textbf {\bibinfo {volume} {18}},\ \bibinfo {pages} {228--235} (\bibinfo
  {year} {2008})}\BibitemShut {NoStop}%
\bibitem [{\citenamefont {Johnson}(1955)}]{johnson1955particle}%
  \BibitemOpen
  \bibfield  {author} {\bibinfo {author} {\bibfnamefont {A.}~\bibnamefont
  {Johnson}},\ }\bibfield  {title} {\enquote {\bibinfo {title} {Particle size
  distribution in clays},}\ }\href@noop {} {\bibfield  {journal} {\bibinfo
  {journal} {Bulletin}\ ,\ \bibinfo {pages} {89}} (\bibinfo {year}
  {1955})}\BibitemShut {NoStop}%
\bibitem [{\citenamefont {Lavrinenko}, \citenamefont {Wohlleben},\ and\
  \citenamefont {Leyrer}(2009)}]{lavrinenko2009influence}%
  \BibitemOpen
  \bibfield  {author} {\bibinfo {author} {\bibfnamefont {A.~V.}\ \bibnamefont
  {Lavrinenko}}, \bibinfo {author} {\bibfnamefont {W.}~\bibnamefont
  {Wohlleben}}, \ and\ \bibinfo {author} {\bibfnamefont {R.~J.}\ \bibnamefont
  {Leyrer}},\ }\bibfield  {title} {\enquote {\bibinfo {title} {Influence of
  imperfections on the photonic insulating and guiding properties of finite
  si-inverted opal crystals},}\ }\href@noop {} {\bibfield  {journal} {\bibinfo
  {journal} {Optics Express}\ }\textbf {\bibinfo {volume} {17}},\ \bibinfo
  {pages} {747--760} (\bibinfo {year} {2009})}\BibitemShut {NoStop}%
\bibitem [{\citenamefont {Li}\ \emph {et~al.}(2003)\citenamefont {Li},
  \citenamefont {Armes}, \citenamefont {Jin},\ and\ \citenamefont
  {Zhu}}]{li2003direct}%
  \BibitemOpen
  \bibfield  {author} {\bibinfo {author} {\bibfnamefont {Y.}~\bibnamefont
  {Li}}, \bibinfo {author} {\bibfnamefont {S.~P.}\ \bibnamefont {Armes}},
  \bibinfo {author} {\bibfnamefont {X.}~\bibnamefont {Jin}}, \ and\ \bibinfo
  {author} {\bibfnamefont {S.}~\bibnamefont {Zhu}},\ }\bibfield  {title}
  {\enquote {\bibinfo {title} {Direct synthesis of well-defined quaternized
  homopolymers and diblock copolymers via atrp in protic media},}\ }\href@noop
  {} {\bibfield  {journal} {\bibinfo  {journal} {Macromolecules}\ }\textbf
  {\bibinfo {volume} {36}},\ \bibinfo {pages} {8268--8275} (\bibinfo {year}
  {2003})}\BibitemShut {NoStop}%
\bibitem [{\citenamefont {Fraden}(1995)}]{fraden1995phase}%
  \BibitemOpen
  \bibfield  {author} {\bibinfo {author} {\bibfnamefont {S.}~\bibnamefont
  {Fraden}},\ }\bibfield  {title} {\enquote {\bibinfo {title} {Phase
  transitions in colloidal suspensions of virus particles},}\ }in\ \href@noop
  {} {\emph {\bibinfo {booktitle} {Observation, Prediction and Simulation of
  Phase Transitions in Complex Fluids}}}\ (\bibinfo  {publisher} {Springer},\
  \bibinfo {year} {1995})\ pp.\ \bibinfo {pages} {113--164}\BibitemShut
  {NoStop}%
\bibitem [{\citenamefont {Gazzillo}\ and\ \citenamefont
  {Giacometti}(2011)}]{gazzillo2011effects}%
  \BibitemOpen
  \bibfield  {author} {\bibinfo {author} {\bibfnamefont {D.}~\bibnamefont
  {Gazzillo}}\ and\ \bibinfo {author} {\bibfnamefont {A.}~\bibnamefont
  {Giacometti}},\ }\bibfield  {title} {\enquote {\bibinfo {title} {Effects of
  polydispersity and anisotropy in colloidal and protein solutions: An integral
  equation approach},}\ }\href@noop {} {\bibfield  {journal} {\bibinfo
  {journal} {Interdisciplinary Sciences: Computational Life Sciences}\ }\textbf
  {\bibinfo {volume} {3}},\ \bibinfo {pages} {251--265} (\bibinfo {year}
  {2011})}\BibitemShut {NoStop}%
\bibitem [{\citenamefont {Vera}\ \emph {et~al.}(1996)\citenamefont {Vera},
  \citenamefont {Gallardo}, \citenamefont {Salcedo},\ and\ \citenamefont
  {Delgado}}]{vera1996colloidal}%
  \BibitemOpen
  \bibfield  {author} {\bibinfo {author} {\bibfnamefont {P.}~\bibnamefont
  {Vera}}, \bibinfo {author} {\bibfnamefont {V.}~\bibnamefont {Gallardo}},
  \bibinfo {author} {\bibfnamefont {J.}~\bibnamefont {Salcedo}}, \ and\
  \bibinfo {author} {\bibfnamefont {A.}~\bibnamefont {Delgado}},\ }\bibfield
  {title} {\enquote {\bibinfo {title} {Colloidal stability of a pharmaceutical
  latex: experimental determinations and theoretical predictions},}\
  }\href@noop {} {\bibfield  {journal} {\bibinfo  {journal} {Journal of Colloid
  and Interface Science}\ }\textbf {\bibinfo {volume} {177}},\ \bibinfo {pages}
  {553--560} (\bibinfo {year} {1996})}\BibitemShut {NoStop}%
\bibitem [{\citenamefont {Park}\ \emph {et~al.}(2009)\citenamefont {Park},
  \citenamefont {Lee}, \citenamefont {Lee}, \citenamefont {Hur},\ and\
  \citenamefont {Park}}]{park2009spectroscopic}%
  \BibitemOpen
  \bibfield  {author} {\bibinfo {author} {\bibfnamefont {M.-H.}\ \bibnamefont
  {Park}}, \bibinfo {author} {\bibfnamefont {T.-H.}\ \bibnamefont {Lee}},
  \bibinfo {author} {\bibfnamefont {B.-M.}\ \bibnamefont {Lee}}, \bibinfo
  {author} {\bibfnamefont {J.}~\bibnamefont {Hur}}, \ and\ \bibinfo {author}
  {\bibfnamefont {D.-H.}\ \bibnamefont {Park}},\ }\bibfield  {title} {\enquote
  {\bibinfo {title} {Spectroscopic and chromatographic characterization of
  wastewater organic matter from a biological treatment plant},}\ }\href@noop
  {} {\bibfield  {journal} {\bibinfo  {journal} {Sensors}\ }\textbf {\bibinfo
  {volume} {10}},\ \bibinfo {pages} {254--265} (\bibinfo {year}
  {2009})}\BibitemShut {NoStop}%
\bibitem [{\citenamefont {Sollich}(2002)}]{PeterReview}%
  \BibitemOpen
  \bibfield  {author} {\bibinfo {author} {\bibfnamefont {P.}~\bibnamefont
  {Sollich}},\ }\bibfield  {title} {\enquote {\bibinfo {title} {Predicting
  phase equilibria in polydisperse systems},}\ }\href
  {http://stacks.iop.org/0953-8984/14/i=3/a=201} {\bibfield  {journal}
  {\bibinfo  {journal} {Journal of Physics: Condensed Matter}\ }\textbf
  {\bibinfo {volume} {14}},\ \bibinfo {pages} {R79} (\bibinfo {year}
  {2002})}\BibitemShut {NoStop}%
\bibitem [{\citenamefont {Weiner}, \citenamefont {Langley},\ and\ \citenamefont
  {Ford}(1974)}]{PhysRevLett.32.879}%
  \BibitemOpen
  \bibfield  {author} {\bibinfo {author} {\bibfnamefont {J.}~\bibnamefont
  {Weiner}}, \bibinfo {author} {\bibfnamefont {K.~H.}\ \bibnamefont {Langley}},
  \ and\ \bibinfo {author} {\bibfnamefont {N.~C.}\ \bibnamefont {Ford}},\
  }\bibfield  {title} {\enquote {\bibinfo {title} {Experimental evidence for a
  departure from the law of the rectilinear diameter},}\ }\href {\doibase
  10.1103/PhysRevLett.32.879} {\bibfield  {journal} {\bibinfo  {journal}
  {Physical Review Letters}\ }\textbf {\bibinfo {volume} {32}},\ \bibinfo
  {pages} {879--881} (\bibinfo {year} {1974})}\BibitemShut {NoStop}%
\bibitem [{\citenamefont {Shimanskaya}\ \emph {et~al.}(1981)\citenamefont
  {Shimanskaya}, \citenamefont {Bezruchko}, \citenamefont {Basok},\ and\
  \citenamefont {Shimanski{\i}}}]{shimanskaya1981experimental}%
  \BibitemOpen
  \bibfield  {author} {\bibinfo {author} {\bibfnamefont {E.}~\bibnamefont
  {Shimanskaya}}, \bibinfo {author} {\bibfnamefont {I.}~\bibnamefont
  {Bezruchko}}, \bibinfo {author} {\bibfnamefont {V.}~\bibnamefont {Basok}}, \
  and\ \bibinfo {author} {\bibfnamefont {Y.}~\bibnamefont {Shimanski{\i}}},\
  }\bibfield  {title} {\enquote {\bibinfo {title} {Experimental determination
  of the critical exponent and of the asymmetric and non asymptotic corrections
  to the equation of the coexistence curve of freon-113},}\ }\href@noop {}
  {\bibfield  {journal} {\bibinfo  {journal} {Soviet Physics--JETP}\ }\textbf
  {\bibinfo {volume} {53}},\ \bibinfo {pages} {139} (\bibinfo {year}
  {1981})}\BibitemShut {NoStop}%
\bibitem [{\citenamefont {Ley-Koo}\ and\ \citenamefont
  {Green}(1977)}]{PhysRevA.16.2483}%
  \BibitemOpen
  \bibfield  {author} {\bibinfo {author} {\bibfnamefont {M.}~\bibnamefont
  {Ley-Koo}}\ and\ \bibinfo {author} {\bibfnamefont {M.~S.}\ \bibnamefont
  {Green}},\ }\bibfield  {title} {\enquote {\bibinfo {title} {Revised and
  extended scaling for coexisting densities of s${\mathrm{f}}_{6}$},}\ }\href
  {\doibase 10.1103/PhysRevA.16.2483} {\bibfield  {journal} {\bibinfo
  {journal} {Physical Review A}\ }\textbf {\bibinfo {volume} {16}},\ \bibinfo
  {pages} {2483--2487} (\bibinfo {year} {1977})}\BibitemShut {NoStop}%
\bibitem [{\citenamefont {Lee}\ and\ \citenamefont
  {Yang}(1952)}]{PhysRev.87.410}%
  \BibitemOpen
  \bibfield  {author} {\bibinfo {author} {\bibfnamefont {T.~D.}\ \bibnamefont
  {Lee}}\ and\ \bibinfo {author} {\bibfnamefont {C.~N.}\ \bibnamefont {Yang}},\
  }\bibfield  {title} {\enquote {\bibinfo {title} {Statistical theory of
  equations of state and phase transitions. ii. lattice gas and ising model},}\
  }\href {\doibase 10.1103/PhysRev.87.410} {\bibfield  {journal} {\bibinfo
  {journal} {Physical Review}\ }\textbf {\bibinfo {volume} {87}},\ \bibinfo
  {pages} {410--419} (\bibinfo {year} {1952})}\BibitemShut {NoStop}%
\bibitem [{\citenamefont {Yang}\ and\ \citenamefont
  {Yang}(1964)}]{PhysRevLett.13.303}%
  \BibitemOpen
  \bibfield  {author} {\bibinfo {author} {\bibfnamefont {C.~N.}\ \bibnamefont
  {Yang}}\ and\ \bibinfo {author} {\bibfnamefont {C.~P.}\ \bibnamefont
  {Yang}},\ }\bibfield  {title} {\enquote {\bibinfo {title} {Critical point in
  liquid-gas transitions},}\ }\href {\doibase 10.1103/PhysRevLett.13.303}
  {\bibfield  {journal} {\bibinfo  {journal} {Physical Review Letters}\
  }\textbf {\bibinfo {volume} {13}},\ \bibinfo {pages} {303--305} (\bibinfo
  {year} {1964})}\BibitemShut {NoStop}%
\bibitem [{\citenamefont {Widom}(1965)}]{widom1965eqnstate}%
  \BibitemOpen
  \bibfield  {author} {\bibinfo {author} {\bibfnamefont {B.}~\bibnamefont
  {Widom}},\ }\bibfield  {title} {\enquote {\bibinfo {title} {Equation of state
  in the neighborhood of the critical point},}\ }\href {\doibase
  10.1063/1.1696618} {\bibfield  {journal} {\bibinfo  {journal} {The Journal of
  Chemical Physics}\ }\textbf {\bibinfo {volume} {43}},\ \bibinfo {pages}
  {3898--3905} (\bibinfo {year} {1965})},\ \Eprint
  {http://arxiv.org/abs/https://doi.org/10.1063/1.1696618}
  {https://doi.org/10.1063/1.1696618} \BibitemShut {NoStop}%
\bibitem [{\citenamefont {Rehr}\ and\ \citenamefont
  {Mermin}(1973)}]{rehr1973revised}%
  \BibitemOpen
  \bibfield  {author} {\bibinfo {author} {\bibfnamefont {J.}~\bibnamefont
  {Rehr}}\ and\ \bibinfo {author} {\bibfnamefont {N.}~\bibnamefont {Mermin}},\
  }\bibfield  {title} {\enquote {\bibinfo {title} {Revised scaling equation of
  state at the liquid-vapor critical point},}\ }\href@noop {} {\bibfield
  {journal} {\bibinfo  {journal} {Physical Review A}\ }\textbf {\bibinfo
  {volume} {8}},\ \bibinfo {pages} {472} (\bibinfo {year} {1973})}\BibitemShut
  {NoStop}%
\bibitem [{\citenamefont {Kim}, \citenamefont {Fisher},\ and\ \citenamefont
  {Orkoulas}(2003)}]{PhysRevE.67.061506}%
  \BibitemOpen
  \bibfield  {author} {\bibinfo {author} {\bibfnamefont {Y.~C.}\ \bibnamefont
  {Kim}}, \bibinfo {author} {\bibfnamefont {M.~E.}\ \bibnamefont {Fisher}}, \
  and\ \bibinfo {author} {\bibfnamefont {G.}~\bibnamefont {Orkoulas}},\
  }\bibfield  {title} {\enquote {\bibinfo {title} {{Asymmetric fluid
  criticality. I. Scaling with pressure mixing}},}\ }\href {\doibase
  10.1103/PhysRevE.67.061506} {\bibfield  {journal} {\bibinfo  {journal}
  {Physical Review E}\ }\textbf {\bibinfo {volume} {67}},\ \bibinfo {pages}
  {61506} (\bibinfo {year} {2003})}\BibitemShut {NoStop}%
\bibitem [{\citenamefont {Anisimov}\ and\ \citenamefont
  {Wang}(2006)}]{PhysRevLett.97.025703}%
  \BibitemOpen
  \bibfield  {author} {\bibinfo {author} {\bibfnamefont {M.~A.}\ \bibnamefont
  {Anisimov}}\ and\ \bibinfo {author} {\bibfnamefont {J.}~\bibnamefont
  {Wang}},\ }\bibfield  {title} {\enquote {\bibinfo {title} {Nature of
  asymmetry in fluid criticality},}\ }\href {\doibase
  10.1103/PhysRevLett.97.025703} {\bibfield  {journal} {\bibinfo  {journal}
  {Physical Review Letters}\ }\textbf {\bibinfo {volume} {97}},\ \bibinfo
  {pages} {025703} (\bibinfo {year} {2006})}\BibitemShut {NoStop}%
\bibitem [{\citenamefont {Pestak}\ \emph {et~al.}(1987)\citenamefont {Pestak},
  \citenamefont {Goldstein}, \citenamefont {Chan}, \citenamefont {de~Bruyn},
  \citenamefont {Balzarini},\ and\ \citenamefont {Ashcroft}}]{PhysRevB.36.599}%
  \BibitemOpen
  \bibfield  {author} {\bibinfo {author} {\bibfnamefont {M.~W.}\ \bibnamefont
  {Pestak}}, \bibinfo {author} {\bibfnamefont {R.~E.}\ \bibnamefont
  {Goldstein}}, \bibinfo {author} {\bibfnamefont {M.~H.~W.}\ \bibnamefont
  {Chan}}, \bibinfo {author} {\bibfnamefont {J.~R.}\ \bibnamefont {de~Bruyn}},
  \bibinfo {author} {\bibfnamefont {D.~A.}\ \bibnamefont {Balzarini}}, \ and\
  \bibinfo {author} {\bibfnamefont {N.~W.}\ \bibnamefont {Ashcroft}},\
  }\bibfield  {title} {\enquote {\bibinfo {title} {Three-body interactions,
  scaling variables, and singular diameters in the coexistence curves of
  fluids},}\ }\href {\doibase 10.1103/PhysRevB.36.599} {\bibfield  {journal}
  {\bibinfo  {journal} {Physical Review B}\ }\textbf {\bibinfo {volume} {36}},\
  \bibinfo {pages} {599--614} (\bibinfo {year} {1987})}\BibitemShut {NoStop}%
\bibitem [{\citenamefont {Sollich}\ and\ \citenamefont
  {Wilding}(2011)}]{sollich2011polydispersity}%
  \BibitemOpen
  \bibfield  {author} {\bibinfo {author} {\bibfnamefont {P.}~\bibnamefont
  {Sollich}}\ and\ \bibinfo {author} {\bibfnamefont {N.~B.}\ \bibnamefont
  {Wilding}},\ }\bibfield  {title} {\enquote {\bibinfo {title} {Polydispersity
  induced solid--solid transitions in model colloids},}\ }\href@noop {}
  {\bibfield  {journal} {\bibinfo  {journal} {Soft Matter}\ }\textbf {\bibinfo
  {volume} {7}},\ \bibinfo {pages} {4472--4484} (\bibinfo {year}
  {2011})}\BibitemShut {NoStop}%
\bibitem [{\citenamefont {Sluckin}(1989)}]{sluckin1989polydispersity}%
  \BibitemOpen
  \bibfield  {author} {\bibinfo {author} {\bibfnamefont {T.}~\bibnamefont
  {Sluckin}},\ }\bibfield  {title} {\enquote {\bibinfo {title} {Polydispersity
  in liquid crystal systems},}\ }\href@noop {} {\bibfield  {journal} {\bibinfo
  {journal} {Liquid Crystals}\ }\textbf {\bibinfo {volume} {6}},\ \bibinfo
  {pages} {111--131} (\bibinfo {year} {1989})}\BibitemShut {NoStop}%
\bibitem [{\citenamefont {Liddle}, \citenamefont {Narayanan},\ and\
  \citenamefont {Poon}(2011)}]{liddle2011polydispersity}%
  \BibitemOpen
  \bibfield  {author} {\bibinfo {author} {\bibfnamefont {S.}~\bibnamefont
  {Liddle}}, \bibinfo {author} {\bibfnamefont {T.}~\bibnamefont {Narayanan}}, \
  and\ \bibinfo {author} {\bibfnamefont {W.}~\bibnamefont {Poon}},\ }\bibfield
  {title} {\enquote {\bibinfo {title} {Polydispersity effects in
  colloid--polymer mixtures},}\ }\href@noop {} {\bibfield  {journal} {\bibinfo
  {journal} {Journal of Physics: Condensed Matter}\ }\textbf {\bibinfo {volume}
  {23}},\ \bibinfo {pages} {194116} (\bibinfo {year} {2011})}\BibitemShut
  {NoStop}%
\bibitem [{\citenamefont {Wang}\ \emph {et~al.}(2008)\citenamefont {Wang},
  \citenamefont {Cerdeirina}, \citenamefont {Anisimov},\ and\ \citenamefont
  {Sengers}}]{ISI:000254539700040}%
  \BibitemOpen
  \bibfield  {author} {\bibinfo {author} {\bibfnamefont {J.}~\bibnamefont
  {Wang}}, \bibinfo {author} {\bibfnamefont {C.~A.}\ \bibnamefont
  {Cerdeirina}}, \bibinfo {author} {\bibfnamefont {M.~A.}\ \bibnamefont
  {Anisimov}}, \ and\ \bibinfo {author} {\bibfnamefont {J.~V.}\ \bibnamefont
  {Sengers}},\ }\bibfield  {title} {\enquote {\bibinfo {title} {{Principle of
  isomorphism and complete scaling for binary-fluid criticality}},}\ }\href
  {\doibase 10.1103/PhysRevE.77.031127} {\bibfield  {journal} {\bibinfo
  {journal} {Physical Review E}\ }\textbf {\bibinfo {volume} {77}} (\bibinfo
  {year} {2008}),\ 10.1103/PhysRevE.77.031127}\BibitemShut {NoStop}%
\bibitem [{\citenamefont {Perez-Sanchez}\ \emph {et~al.}(2010)\citenamefont
  {Perez-Sanchez}, \citenamefont {Losada-Perez}, \citenamefont {Cerdeirina},
  \citenamefont {Sengers},\ and\ \citenamefont
  {Anisimov}}]{ISI:000276971500024}%
  \BibitemOpen
  \bibfield  {author} {\bibinfo {author} {\bibfnamefont {G.}~\bibnamefont
  {Perez-Sanchez}}, \bibinfo {author} {\bibfnamefont {P.}~\bibnamefont
  {Losada-Perez}}, \bibinfo {author} {\bibfnamefont {C.~A.}\ \bibnamefont
  {Cerdeirina}}, \bibinfo {author} {\bibfnamefont {J.~V.}\ \bibnamefont
  {Sengers}}, \ and\ \bibinfo {author} {\bibfnamefont {M.~A.}\ \bibnamefont
  {Anisimov}},\ }\bibfield  {title} {\enquote {\bibinfo {title} {{Asymmetric
  criticality in weakly compressible liquid mixtures}},}\ }\href {\doibase
  10.1063/1.3378626} {\bibfield  {journal} {\bibinfo  {journal} {The Journal of
  Chemical Physics}\ }\textbf {\bibinfo {volume} {132}} (\bibinfo {year}
  {2010}),\ 10.1063/1.3378626}\BibitemShut {NoStop}%
\bibitem [{\citenamefont {Belyakov}\ \emph {et~al.}(2013)\citenamefont
  {Belyakov}, \citenamefont {Gorodetskii}, \citenamefont {Kulikov},
  \citenamefont {Muratov}, \citenamefont {Voronov}, \citenamefont {Grigoriev},\
  and\ \citenamefont {Volkov}}]{Belyakov2013}%
  \BibitemOpen
  \bibfield  {author} {\bibinfo {author} {\bibfnamefont {M.}~\bibnamefont
  {Belyakov}}, \bibinfo {author} {\bibfnamefont {E.}~\bibnamefont
  {Gorodetskii}}, \bibinfo {author} {\bibfnamefont {V.}~\bibnamefont
  {Kulikov}}, \bibinfo {author} {\bibfnamefont {A.}~\bibnamefont {Muratov}},
  \bibinfo {author} {\bibfnamefont {V.}~\bibnamefont {Voronov}}, \bibinfo
  {author} {\bibfnamefont {B.}~\bibnamefont {Grigoriev}}, \ and\ \bibinfo
  {author} {\bibfnamefont {A.}~\bibnamefont {Volkov}},\ }\bibfield  {title}
  {\enquote {\bibinfo {title} {{Anomalous properties of dew-bubble curves in
  the vicinity of liquid–vapor critical points}},}\ }\href {\doibase
  10.1016/j.fluid.2013.07.040} {\bibfield  {journal} {\bibinfo  {journal}
  {Fluid Phase Equilibria}\ }\textbf {\bibinfo {volume} {358}},\ \bibinfo
  {pages} {91--97} (\bibinfo {year} {2013})}\BibitemShut {NoStop}%
\bibitem [{\citenamefont {Belyakov}\ \emph {et~al.}(2014)\citenamefont
  {Belyakov}, \citenamefont {Gorodetskii}, \citenamefont {Kulikov},
  \citenamefont {Voronov},\ and\ \citenamefont
  {Grigoriev}}]{ISI:000345509900008}%
  \BibitemOpen
  \bibfield  {author} {\bibinfo {author} {\bibfnamefont {M.~Y.}\ \bibnamefont
  {Belyakov}}, \bibinfo {author} {\bibfnamefont {E.~E.}\ \bibnamefont
  {Gorodetskii}}, \bibinfo {author} {\bibfnamefont {V.~D.}\ \bibnamefont
  {Kulikov}}, \bibinfo {author} {\bibfnamefont {V.~P.}\ \bibnamefont
  {Voronov}}, \ and\ \bibinfo {author} {\bibfnamefont {B.~A.}\ \bibnamefont
  {Grigoriev}},\ }\bibfield  {title} {\enquote {\bibinfo {title} {{Scaled
  equation of state for multi-component fluids}},}\ }\href {\doibase
  10.1016/j.chemphys.2014.10.009} {\bibfield  {journal} {\bibinfo  {journal}
  {Chemical Physics}\ }\textbf {\bibinfo {volume} {445}},\ \bibinfo {pages}
  {53--58} (\bibinfo {year} {2014})}\BibitemShut {NoStop}%
\bibitem [{\citenamefont {Wilding}\ \emph {et~al.}(2006)\citenamefont
  {Wilding}, \citenamefont {Sollich}, \citenamefont {Fasolo},\ and\
  \citenamefont {Buzzacchi}}]{LJNigelPeter}%
  \BibitemOpen
  \bibfield  {author} {\bibinfo {author} {\bibfnamefont {N.~B.}\ \bibnamefont
  {Wilding}}, \bibinfo {author} {\bibfnamefont {P.}~\bibnamefont {Sollich}},
  \bibinfo {author} {\bibfnamefont {M.}~\bibnamefont {Fasolo}}, \ and\ \bibinfo
  {author} {\bibfnamefont {M.}~\bibnamefont {Buzzacchi}},\ }\bibfield  {title}
  {\enquote {\bibinfo {title} {Phase behavior and particle size cutoff effects
  in polydisperse fluids},}\ }\href {\doibase 10.1063/1.2208358} {\bibfield
  {journal} {\bibinfo  {journal} {The Journal of Chemical Physics}\ }\textbf
  {\bibinfo {volume} {125}},\ \bibinfo {pages} {014908} (\bibinfo {year}
  {2006})},\ \Eprint {http://arxiv.org/abs/https://doi.org/10.1063/1.2208358}
  {https://doi.org/10.1063/1.2208358} \BibitemShut {NoStop}%
\bibitem [{\citenamefont {Ozawa}\ and\ \citenamefont
  {Berthier}(2017)}]{BerthierEntropyPoly}%
  \BibitemOpen
  \bibfield  {author} {\bibinfo {author} {\bibfnamefont {M.}~\bibnamefont
  {Ozawa}}\ and\ \bibinfo {author} {\bibfnamefont {L.}~\bibnamefont
  {Berthier}},\ }\bibfield  {title} {\enquote {\bibinfo {title} {Does the
  configurational entropy of polydisperse particles exist?}}\ }\href {\doibase
  10.1063/1.4972525} {\bibfield  {journal} {\bibinfo  {journal} {The Journal of
  Chemical Physics}\ }\textbf {\bibinfo {volume} {146}},\ \bibinfo {pages}
  {014502} (\bibinfo {year} {2017})},\ \Eprint
  {http://arxiv.org/abs/https://doi.org/10.1063/1.4972525}
  {https://doi.org/10.1063/1.4972525} \BibitemShut {NoStop}%
\bibitem [{\citenamefont {Schofield}, \citenamefont {Litster},\ and\
  \citenamefont {Ho}(1969)}]{schofield1969correlation}%
  \BibitemOpen
  \bibfield  {author} {\bibinfo {author} {\bibfnamefont {P.}~\bibnamefont
  {Schofield}}, \bibinfo {author} {\bibfnamefont {J.}~\bibnamefont {Litster}},
  \ and\ \bibinfo {author} {\bibfnamefont {J.~T.}\ \bibnamefont {Ho}},\
  }\bibfield  {title} {\enquote {\bibinfo {title} {Correlation between critical
  coefficients and critical exponents},}\ }\href@noop {} {\bibfield  {journal}
  {\bibinfo  {journal} {Physical Review Letters}\ }\textbf {\bibinfo {volume}
  {23}},\ \bibinfo {pages} {1098} (\bibinfo {year} {1969})}\BibitemShut
  {NoStop}%
\bibitem [{\citenamefont {Sanchez}, \citenamefont {Meichle},\ and\
  \citenamefont {Garland}(1983)}]{PhysRevA.28.1647}%
  \BibitemOpen
  \bibfield  {author} {\bibinfo {author} {\bibfnamefont {G.}~\bibnamefont
  {Sanchez}}, \bibinfo {author} {\bibfnamefont {M.}~\bibnamefont {Meichle}}, \
  and\ \bibinfo {author} {\bibfnamefont {C.~W.}\ \bibnamefont {Garland}},\
  }\bibfield  {title} {\enquote {\bibinfo {title} {Critical heat capacity in a
  3-methylpentane + nitroethane mixture near its consolute point},}\ }\href
  {\doibase 10.1103/PhysRevA.28.1647} {\bibfield  {journal} {\bibinfo
  {journal} {Physical Review A}\ }\textbf {\bibinfo {volume} {28}},\ \bibinfo
  {pages} {1647--1653} (\bibinfo {year} {1983})}\BibitemShut {NoStop}%
\bibitem [{\citenamefont {Fisher}(1968)}]{PhysRev.176.257}%
  \BibitemOpen
  \bibfield  {author} {\bibinfo {author} {\bibfnamefont {M.~E.}\ \bibnamefont
  {Fisher}},\ }\bibfield  {title} {\enquote {\bibinfo {title} {{Renormalization
  of Critical Exponents by Hidden Variables}},}\ }\href {\doibase
  10.1103/PhysRev.176.257} {\bibfield  {journal} {\bibinfo  {journal} {Physical
  Review}\ }\textbf {\bibinfo {volume} {176}},\ \bibinfo {pages} {257--272}
  (\bibinfo {year} {1968})}\BibitemShut {NoStop}%
\bibitem [{\citenamefont {Anisimov}\ \emph {et~al.}(1995)\citenamefont
  {Anisimov}, \citenamefont {Gorodetskii}, \citenamefont {Kulikov},
  \citenamefont {Povodyrev},\ and\ \citenamefont {Sengers}}]{Anisimov1995}%
  \BibitemOpen
  \bibfield  {author} {\bibinfo {author} {\bibfnamefont {M.}~\bibnamefont
  {Anisimov}}, \bibinfo {author} {\bibfnamefont {E.}~\bibnamefont
  {Gorodetskii}}, \bibinfo {author} {\bibfnamefont {V.}~\bibnamefont
  {Kulikov}}, \bibinfo {author} {\bibfnamefont {A.}~\bibnamefont {Povodyrev}},
  \ and\ \bibinfo {author} {\bibfnamefont {J.}~\bibnamefont {Sengers}},\
  }\bibfield  {title} {\enquote {\bibinfo {title} {{A general isomorphism
  approach to thermodynamic and transport properties of binary fluid mixtures
  near critical points}},}\ }\href {\doibase 10.1016/0378-4371(95)00217-U}
  {\bibfield  {journal} {\bibinfo  {journal} {Physica A: Statistical Mechanics
  and its Applications}\ }\textbf {\bibinfo {volume} {220}},\ \bibinfo {pages}
  {277--324} (\bibinfo {year} {1995})}\BibitemShut {NoStop}%
\bibitem [{\citenamefont {Anisimov}(2007)}]{ISI:000243587200033}%
  \BibitemOpen
  \bibfield  {author} {\bibinfo {author} {\bibfnamefont {M.~A.}\ \bibnamefont
  {Anisimov}},\ }\bibfield  {title} {\enquote {\bibinfo {title} {{Divergence of
  Tolman's length for a droplet near the critical point}},}\ }\href {\doibase
  10.1103/PhysRevLett.98.035702} {\bibfield  {journal} {\bibinfo  {journal}
  {Physical Review Letters}\ }\textbf {\bibinfo {volume} {98}} (\bibinfo {year}
  {2007}),\ 10.1103/PhysRevLett.98.035702}\BibitemShut {NoStop}%
\bibitem [{\citenamefont {Losada-Perez}\ \emph
  {et~al.}(2010{\natexlab{a}})\citenamefont {Losada-Perez}, \citenamefont
  {Perez-Sanchez}, \citenamefont {Cerdeirina},\ and\ \citenamefont
  {Thoen}}]{ISI:000277265700029}%
  \BibitemOpen
  \bibfield  {author} {\bibinfo {author} {\bibfnamefont {P.}~\bibnamefont
  {Losada-Perez}}, \bibinfo {author} {\bibfnamefont {G.}~\bibnamefont
  {Perez-Sanchez}}, \bibinfo {author} {\bibfnamefont {C.~A.}\ \bibnamefont
  {Cerdeirina}}, \ and\ \bibinfo {author} {\bibfnamefont {J.}~\bibnamefont
  {Thoen}},\ }\bibfield  {title} {\enquote {\bibinfo {title} {{Dielectric
  constant of fluids and fluid mixtures at criticality}},}\ }\href {\doibase
  10.1103/PhysRevE.81.041121} {\bibfield  {journal} {\bibinfo  {journal}
  {Physical Review E}\ }\textbf {\bibinfo {volume} {81}} (\bibinfo {year}
  {2010}{\natexlab{a}}),\ 10.1103/PhysRevE.81.041121}\BibitemShut {NoStop}%
\bibitem [{\citenamefont {Losada-P{\'{e}}rez}, \citenamefont {Glorieux},\ and\
  \citenamefont {Thoen}(2012)}]{losada2012critical}%
  \BibitemOpen
  \bibfield  {author} {\bibinfo {author} {\bibfnamefont {P.}~\bibnamefont
  {Losada-P{\'{e}}rez}}, \bibinfo {author} {\bibfnamefont {C.}~\bibnamefont
  {Glorieux}}, \ and\ \bibinfo {author} {\bibfnamefont {J.}~\bibnamefont
  {Thoen}},\ }\bibfield  {title} {\enquote {\bibinfo {title} {{The critical
  behavior of the refractive index near liquid-liquid critical points}},}\
  }\href@noop {} {\bibfield  {journal} {\bibinfo  {journal} {The Journal of
  Chemical Physics}\ }\textbf {\bibinfo {volume} {136}},\ \bibinfo {pages}
  {144502} (\bibinfo {year} {2012})}\BibitemShut {NoStop}%
\bibitem [{\citenamefont {Losada-Perez}\ \emph
  {et~al.}(2010{\natexlab{b}})\citenamefont {Losada-Perez}, \citenamefont
  {Perez-Sanchez}, \citenamefont {Troncoso},\ and\ \citenamefont
  {Cerdeirina}}]{ISI:000276971500031}%
  \BibitemOpen
  \bibfield  {author} {\bibinfo {author} {\bibfnamefont {P.}~\bibnamefont
  {Losada-Perez}}, \bibinfo {author} {\bibfnamefont {G.}~\bibnamefont
  {Perez-Sanchez}}, \bibinfo {author} {\bibfnamefont {J.}~\bibnamefont
  {Troncoso}}, \ and\ \bibinfo {author} {\bibfnamefont {C.~A.}\ \bibnamefont
  {Cerdeirina}},\ }\bibfield  {title} {\enquote {\bibinfo {title} {{Heat
  capacity anomalies along the critical isotherm in fluid-fluid phase
  transitions}},}\ }\href {\doibase 10.1063/1.3374819} {\bibfield  {journal}
  {\bibinfo  {journal} {The Journal of Chemical Physics}\ }\textbf {\bibinfo
  {volume} {132}} (\bibinfo {year} {2010}{\natexlab{b}}),\
  10.1063/1.3374819}\BibitemShut {NoStop}%
\bibitem [{\citenamefont {Jamali}\ and\ \citenamefont
  {Behnejad}(2014)}]{ISI:000337604100004}%
  \BibitemOpen
  \bibfield  {author} {\bibinfo {author} {\bibfnamefont {A.}~\bibnamefont
  {Jamali}}\ and\ \bibinfo {author} {\bibfnamefont {H.}~\bibnamefont
  {Behnejad}},\ }\bibfield  {title} {\enquote {\bibinfo {title} {{Critical
  behaviour of thermo-physical properties in weakly compressible liquid
  mixtures}},}\ }\href {\doibase 10.1080/00319104.2013.842471} {\bibfield
  {journal} {\bibinfo  {journal} {Physics and Chemistry of Liquids}\ }\textbf
  {\bibinfo {volume} {52}},\ \bibinfo {pages} {519--532} (\bibinfo {year}
  {2014})}\BibitemShut {NoStop}%
\bibitem [{\citenamefont {Behnejad}, \citenamefont {Cheshmpak},\ and\
  \citenamefont {Jamali}(2015)}]{behnejad2015isomorphic}%
  \BibitemOpen
  \bibfield  {author} {\bibinfo {author} {\bibfnamefont {H.}~\bibnamefont
  {Behnejad}}, \bibinfo {author} {\bibfnamefont {H.}~\bibnamefont {Cheshmpak}},
  \ and\ \bibinfo {author} {\bibfnamefont {A.}~\bibnamefont {Jamali}},\
  }\bibfield  {title} {\enquote {\bibinfo {title} {Isomorphic viscosity
  equation of state for binary fluid mixtures},}\ }\href@noop {} {\bibfield
  {journal} {\bibinfo  {journal} {Acta Chimica Slovenica}\ }\textbf {\bibinfo
  {volume} {62}},\ \bibinfo {pages} {754--760} (\bibinfo {year}
  {2015})}\BibitemShut {NoStop}%
\bibitem [{\citenamefont {Sollich}(2008)}]{sollich2008weakly}%
  \BibitemOpen
  \bibfield  {author} {\bibinfo {author} {\bibfnamefont {P.}~\bibnamefont
  {Sollich}},\ }\bibfield  {title} {\enquote {\bibinfo {title} {Weakly
  polydisperse systems: Perturbative phase diagrams that include the critical
  region},}\ }\href@noop {} {\bibfield  {journal} {\bibinfo  {journal}
  {Physical Review Letters}\ }\textbf {\bibinfo {volume} {100}},\ \bibinfo
  {pages} {035701} (\bibinfo {year} {2008})}\BibitemShut {NoStop}%
\end{thebibliography}%

\end{document}